\documentclass[11pt,preprint,flushrt]{aastex}
\usepackage{graphicx,amsmath,subfigure,xcolor}
\usepackage{hyperref} 
\hypersetup{
    colorlinks=true, 
    linktoc=all,     
    linkcolor=blue,  
}
\def\eqq#1{Equation~(\ref{#1})}
\def\etal{{\it et al.}}
\def\ie{{\it i.e.}}

\usepackage[mathlines]{lineno}

\newcommand{\vOmega}{\mbox{\boldmath $\Omega$}}
\newcommand{\vpi}{\mbox{\boldmath $\pi$}}

\newcommand{\vx}{\mbox{${\bf x}$}}

\newcommand{\matA}{\mbox{$A$}}
\newcommand{\vpsf}{\mbox{${\bf PSF}$}}
\newcommand{\vrr}{\mbox{${\bf r}$}}

\newcommand{\vC}{\mbox{${\bf C}$}}
\newcommand{\vc}{\mbox{${\bf c}$}}
\newcommand{\vRate}{\mbox{${\bf Rate}$}}
\newcommand{\vDome}{\mbox{${\bf Dome}$}}
\newcommand{\vSFlat}{\mbox{${\bf SFlat}$}}

\newcommand{\photofit}{\textsc{PhotoFit}}
\newcommand{\wcsfit}{\textsc{WcsFit}}
\newcommand{\scamp}{\textsc{scamp}}
\newcommand{\sextractor}{\textsc{SExtractor}}

\begin{document}
\slugcomment{Version 1.2, submission to PASP, 30 Oct 2017}

\title{Photometric characterization of the Dark Energy Camera}


\def\andname{}

\author{
G.~M.~Bernstein\altaffilmark{1},
T.~M.~C.~Abbott\altaffilmark{2},
R.~Armstrong\altaffilmark{3},
D.~L.~Burke\altaffilmark{4,5},
H.~T.~Diehl\altaffilmark{6},
R.~A.~Gruendl\altaffilmark{7,8},
M.~D.~Johnson\altaffilmark{8},
T.~S.~Li\altaffilmark{6},
E.~S.~Rykoff\altaffilmark{4,5},
A.~R.~Walker\altaffilmark{2},
W.~Wester\altaffilmark{6},
B.~Yanny\altaffilmark{6}
}

\altaffiltext{1}{Department of Physics and Astronomy, University of Pennsylvania, Philadelphia, PA 19104, USA}
\altaffiltext{2}{Cerro Tololo Inter-American Observatory, National Optical Astronomy Observatory, Casilla 603, La Serena, Chile}
\altaffiltext{3}{Department of Astrophysical Sciences, Princeton University, Peyton Hall, Princeton, NJ 08544, USA}
\altaffiltext{4}{Kavli Institute for Particle Astrophysics \& Cosmology, P. O. Box 2450, Stanford University, Stanford, CA 94305, USA}
\altaffiltext{5}{SLAC National Accelerator Laboratory, Menlo Park, CA 94025, USA}
\altaffiltext{6}{Fermi National Accelerator Laboratory, P. O. Box 500, Batavia, IL 60510, USA}
\altaffiltext{7}{Department of Astronomy, University of Illinois, 1002 W. Green Street, Urbana, IL 61801, USA}
\altaffiltext{8}{National Center for Supercomputing Applications, 1205 West Clark St., Urbana, IL 61801, USA}

\begin{abstract}
We characterize the variation in photometric response of the Dark
Energy Camera (DECam) across its 520~Mpix science array during 4
years of operation.  These variations are measured using high
signal-to-noise aperture photometry of $>10^7$ stellar images
in thousands of exposures of a few selected fields, with the
telescope dithered to move the sources around the array.  A
calibration procedure based on these results brings the RMS variation
in aperture magnitudes of bright stars on cloudless nights down to
2--3~mmag, with $<1$~mmag 
of correlated photometric errors for stars separated by
$\ge20\arcsec.$  On cloudless nights, 
any departures of the exposure zeropoints from a secant airmass law
exceeding $>1$~mmag are 
plausibly attributable to spatial/temporal variations in aperture
corrections. These variations can be inferred and corrected
by measuring the fraction of stellar light in an annulus between
6\arcsec\ and 8\arcsec\ diameter.
Key elements of this calibration include: correction
of amplifier nonlinearities; distinguishing pixel-area variations and
stray light from
quantum-efficiency variations in the flat fields; field-dependent
color corrections; and the use of an aperture-correction proxy.
The DECam response pattern across the 2\arcdeg\ field drifts over months by up
to $\pm7$~mmag, in a
nearly-wavelength-independent low-order pattern.  We find no
fundamental barriers to pushing global photometric calibrations toward
mmag accuracy.
\end{abstract}

\section{Introduction}
Photometric calibration of sources in an astronomical imaging survey requires
ensuring that the flux assigned to a given (non-variable) source would
be the same regardless of where on the detector array it is imaged,
where that source lies on the sky, or when the exposure was taken.
The success of the resultant calibration therefore depends on our
ability to model the response of the detector/optics/atmosphere
combination across the focal plane, and the nature and time scales of
variations in this response.  In this work, we perform these
characterizations for the Dark Energy Camera \citep[DECam]{decam}, a
CCD imager with 520 megapixel science array in operation on the 4-meter Blanco telescope
since late 2012.  Increasing precision and homogeneity
of photometric calibration leads to better science results across a
broad range of astrophysical topics pursued by the Dark Energy Survey (DES)
and other projects using DECam: stellar populations and dust
distributions in the Milky Way and its satellites; high-accuracy
galaxy clustering and photometric redshift measurements; studies of
stellar and quasar variability over the 5-year time span of DES; and
accurate Hubble diagrams of supernova flux vs redshift.

We concentrate here on relative photometric calibration, i.e. placing
all of the camera's observations on a consistent flux scale.
Absolute calibration, i.e. determination of the normalization of this
unified flux scale, requires observations of some celestial or
terrestrial sources of known flux, and we defer this discussion to
a future publication.
The traditional approach to relative photometry in
the visible/near-IR, from the days of single-channel photometers, was
to interleave observations of ``standard'' stars of known flux with
observations of targets.  The standards are used
to establish the parameters of a parametric atmosphere$+$instrument
response model for count rates vs source flux, and then invert this
relation to establish fluxes for the targets.  The success of this
approach (as well as the ability to establish the standard-star network to
begin with) rests on a critical assumption, namely: \emph{the response
(including atmospheric transmission) is invariant, to the desired
photometric precision, on the time scale of
the observations used to establish and use its parametric model.}  We
will refer to this assumption of slowly-varying atmospheric
transparency (and instrument response) as the ``atmospheric prior'' on
our photometric calibration.
This assumption is easily violated on nights with clouds.
In this paper we will only examine data taken on cloudless,
a.k.a. ``photometric'' nights, as indicated by the absence of any
clouds on images from the RASICAM thermal-IR all-sky monitor \citep{rasicam}.
See \citet{Burke} for discussion of the spatial and temporal structure of cloud
extinction.

With the advent of large-format array detectors and the ability to perform
quantitative imaging for large numbers of stars simultaneously, the
photometric calibration task gains a critical new responsibility and a
critical new tool.  The responsibility is to determine the relative
response of pixels across the array.  Since the earliest days of CCD
astronomy this has been done by imaging a near-Lambertian source
(either sky background or a dome screen) to generate a ``dome flat''
image that is assumed to reflect the relative response to stellar
illumination across the array.  As detailed in \citet{detrend} and
summarized below, the dome flat technique contains several flaws that
preclude its use for deriving the true photometric response of the
array at percent-level precision (10 mmag) or better.  Fortunately,
the new tool offered by array observations is internal consistency:
the constraint that repeat observations of a given source should yield
the same flux.  Simple applications of the internal consistency
constraint to derive a parametric model of array response involved
scanning a single source across the array \citep{manfroid,brian}.
The first application to survey-scale data, using many stellar images per
exposure, was on the Sloan Digital Sky Survey (SDSS), by \citet{nikhil}.
The DES and the future Large Scale Synoptic
Survey (LSST) choose survey strategies designed to produce
exposures which are highly interlaced in sky position and in time, to better
stabilize photometric solutions based on internal consistency.
This success rests on the second critical assumption of
photometric calibrations, namely that \emph{most individual stars have
  invariant flux, and the median variation in flux of a collection of
  stars is zero.}

It remains true, however, that internal consistency alone is
insufficient to fully calibrate a survey.  Low-frequency modes of
calibration error produce 
weak magnitude differences between overlapping (i.e. nearby)
exposures, which are hence hard to detect and suppress with internal
constraints.  The atmospheric prior ties down large-scale modes if the
survey strategy produces exposures spanning large angles within time
intervals shorter than the stability period.  Thus the best
photometric calibration will combine internal consistency constraints
with an atmospheric prior.  In this paper we answer several questions
that are prerequisite to executing such a calibration of the full
survey:
\begin{enumerate}
\item What parametric model(s) suffice to homogenize the stellar
  photometry across the DECam field of view (FOV) in a single
  exposure?
\item How stable is the atmospheric transmission within a single
cloudless night?  What physical processes are responsible for any
variation and how well can they be modeled and corrected?
\item How does the instrumental response change over intervals between
  24 hours and years? What models and parameters are needed to track
  any variations?
\end{enumerate}
With the answers to these questions in hand, we can proceed to design
the optimal calibration system for DES (and for other observations
made with DECam), which should inform calibration procedures for other
present and future large-FOV imagers.  

Photometric calibration of the first-year DES observations is
described by \citet{y1gold}.  This used the standard technique of
nightly zeropoint/color/extinction solutions using a standard-star
network, incorporating early 
versions of the ``star flats'' described in this paper to flatten
individual exposures, and assigning zeropoints to non-photometric
exposures by forcing agreement with exposures taken in clear
conditions.  The RMS magnitude difference between distinct exposures of a
given region is $\approx3$~mmag, after averaging all the stars in a
CCD-sized region ($\approx10\arcmin$).  RMS differences between this
DES Y1 calibration and magnitudes constructed from APASS \citet{apass}
and 2MASS \citet{twomass} is $\approx20$~mmag, which serves as an
upper limit to the errors in the DES Y1 calibration.  Using the first
three years of DES observations, 
\citet{fgcm} present a first foray into joint
atmospheric/instrumental photometric calibration,
using some of the star-flat calibrations described in this paper
and a physical model for atmospheric extinction.
Reproducibility of stellar photometry at 5--6~mmag RMS is achieved, 
and uniformity of the calibration across the 5000~deg$^2$ DES
footprint is estimated at
7~mmag.  \citet{hypercal} present a cross-calibration of the SDSS
and Pan-Starrs1 (PS1) surveys yielding RMS deviations between the two
at similar level.  This indicates the current state of the art in
photometric calibration of large-scale ground-based sky surveys.  Our
goal is to understand the photometric behavior of DECam at $\approx
1$~mmag RMS level, in hopes of approaching this level of global
calibration accuracy for individual stars in future reductions.  We will therefore ignore
effects which are expected to cause photometric inaccuracies below
this level.

Section~\ref{theory} first recaps the derivations in \citet{detrend} of
the operations necessary to convert raw detector outputs into
homogeneous stellar flux estimates.  Then we describe the formulae and
algorithms used to derive the best parametric model of DECam
instrumental response.
Section~\ref{model} describes the data used to derive the DECam
photometric model, the code used to process it,
and the best-fitting models for DECam
instrumental response,  and then evaluates their performance in homogenizing
short (1-hour) stretches of exposures.  Section~\ref{atmosphere}
addresses the question of just how stable a photometric night is, and
Section~\ref{stability} examines the longer-term variation in DECam
response.  Section~\ref{summary} evaluates the overall level of our
understanding of DECam photometric response, and its implications for
calibration accuracy of DES and other contemporary ground-based
visible sky surveys.

\section{Deriving a response model}
\label{theory}
\subsection{Array response and star flats}
The raw digital values produced by DECam for the signal at pixel
location $x$ in an exposure labeled by index $t$ need to undergo
several detrending steps before we can extract an estimate of the 
top-of-the-atmosphere flux
$f$ of some star in the image.  \citet{detrend} detail the steps
needed to transform the raw data into an image ${\rm Rate}_\star({\bf
  x},t)$ giving 
the rate of photocarrier production in pixel ${\bf x}$ by celestial objects
during exposure $t$.  We will adopt the notation of \citet{detrend}
whereby \textbf{boldface} quantities are vectors over an implied pixel
argument ${\bf x},$ and all mathematical operations are assumed to be
element-wise over pixels.
The steps in the transformation of raw camera data into $\vRate_\star(t)$
include debiasing, conversion from ADU to
photocarriers, linearization of amplifier response, correction of the
``brighter-fatter effect'' \citep{antilogus,gruen}, and background
subtraction.  We refer to \citet{detrend} for descriptions of these
detrending steps, though the linearization and background-subtraction
steps will enter into our analysis of the accuracy of our DECam
photometric model.

A single star with flux $f$ and spectral shape $F_p(\lambda)$
defined by some parameter(s) $p$
produce photocarriers at the rate
\begin{equation}
\vRate_\star(t) = \frac{\Omega_0}{f_1} \vOmega 
\int d\lambda \, r_{\rm ref}(\lambda) \vrr(\lambda,t) \frac{f
  F_p(\lambda) \vpsf(\lambda,t)}{\Omega_0}.
\end{equation}
We define:
\begin{itemize}
\item $\Omega_0$ as the nominal solid angle of an array pixel, and
  \vOmega\ as the fractional deviation of each pixel's size from this value, assumed
  to be independent of wavelength $\lambda$ across a given filter
  band.
\item $r_{\rm ref}(\lambda)$ as the \emph{reference response,} the
  spectral response of a typical 
  pixel in typical atmospheric conditions, otherwise known as the
  ``natural passband'' for the chosen filter.
\item $\vrr(\lambda,t)$ as the \emph{spectral response function}
  giving the true response at a given array position and time relative
  to the reference response.
\item The fraction under the integral gives the apparent surface
  brightness produced by the star, which is defined by the point
  spread function $\vpsf(\lambda, t)$ at the location and time of the
  exposure of the star. 
\item $f_1$ as the flux of a star that produces 1 charge per second in
  nominal conditions.  We will not be concerned with determination of
  this constant, which sets the absolute calibration of the photometry.
\end{itemize}
We opt to normalize these quantities such that
\begin{align}
\sum_x \vpsf(\lambda, t) \vOmega & = 1, \\
\int d\lambda\, F_p(\lambda) r_{\rm ref}(\lambda) & = 1.
\end{align}
The units of $F_p$ are (wavelength)$^{-1}$, and the stellar flux $f$
is then in units of power per unit area.
We will also assume that the PSF is wavelength-independent within any
particular filter band.

Extraction of the stellar flux from the \vRate\ image begins with
defining a \emph{reference spectrum} $F_{\rm ref}(\lambda)$, from
which we can define the \emph{reference flat} field
\begin{equation}
\label{rflat}
\vrr(t) \equiv \int d\lambda\, F_{\rm ref}(\lambda) \vrr(\lambda, t).
\end{equation}
With these definitions, one can show that an estimator for the flux
$f_{\alpha t}$ for star $\alpha$ in exposure $t$ can be produced by
summing over the pixels $x$ in an aperture around the star:
\begin{align}
\label{flux1}
\sum_{x\in \alpha} \frac{\vRate_\star(t)}{\vrr(t)} & =
      \frac{f_{\alpha t}}{f_1} \vC(t;p) \eta(t), \\
\vC(t;p) & \equiv \frac{ \int d\lambda\, F_p(\lambda) r_{\rm
           ref}(\lambda) \vrr(\lambda,t)}{\int d\lambda\, F_{\rm
           ref}(\lambda) r_{\rm ref}(\lambda) \vrr(\lambda,t)} \\
\eta(t) & = \sum_{x \in \alpha} \vpsf(\lambda,t) \vOmega \le 1.
\end{align}
The color correction $\vC(t;p)$ is
evaluated at the focal-plane position of the star and assumed
to be constant across PSF width.  Note
that the color correction is doubly differential, in the sense that it
is unity if either the pixel response is nominal
[$\vrr(\lambda,t)=1$] or the source spectrum matches the reference
[$F_p(\lambda)=F_{\rm ref}(\lambda)$].  
For stellar spectra, we will make a linear approximation in a single
color parameter
\begin{align}
2.5 \log_{10} \vC(t;p) & = p\times \vc(t) \\
p & \equiv  g-i - (g-i)_{\rm ref}.
\end{align}
Spectral synthesis modeling by
\citet{ting} shows that this is accurate to within 1~mmag for stellar
spectra with colors $-1<g-i<1.8$, for the range of spectral response variation
generated by variations in the DECam instrument and atmospheric
variations.
We will only make use of stars in this color range in this paper.
\citet{fgcm} and \citet{detrend} explain how ancillary DECam data and
atmospheric modelling can be used to estimate color corrections for
sources with more exotic spectra.  For our DECam work we take the
reference spectrum to be that of the F8IV star C26202 from the HST
CalSpec standards,\footnote{http://www.stsci.edu/hst/observatory/crds/calspec.html} with
color $(g-i)_{\rm ref}=0.44$ in the natural DECam system.  This
well-characterized star is frequently observed by DES, is near the
median stellar color, and is within DECam's dynamic range.

We assume for now that the fraction of light $\eta$ falling within the
photometric aperture is constant across the FOV for a given exposure,
and in any case this term could be absorbed into the definition of
$\vrr(t)$ if desired. Placing \eqq{flux1} into the magnitude system
yields
\begin{align}
\label{mag1}
m_{\alpha} & = m_1 + m_{\rm inst}(\alpha, t) + p\,\vc(t) + Ap(t), \\
m_{\rm inst}(\alpha,t) & \equiv -2.5 \log_{10} \sum_{x \in \alpha}
                         \frac{\vRate_\star(t)}{\vrr(t)} \\
Ap(t) & \equiv 2.5\log_{10} \eta(t).
\end{align}
The absolute zeropoint $m_1$ is a combination of the instrument flux
normalization $f_1$ and the definition of $m=0$ for the chosen magnitude
system.  The methods described in this paper do not constrain $m_1$.

This analysis shows that the reference flat defined by \eqq{rflat} is
the quantity we want to divide into the image $\vRate_\star$ if we
want to homogenize stellar photometry across the FOV and (in
combination with the aperture correction $Ap$) across time.  For sources
that depart from the reference spectrum, we additionally need to
characterize the color term $\vc(t)$ as it varies across the FOV and
over time.

Traditionally the spatial structure of the reference flat $\vrr(t)$
has been estimated by an image of scene of nearly uniform surface
brightness, such as twilight sky, median night sky, or an illuminated
screen in the dome.  We refer to such an image generically as \vDome.
Perhaps the most important point we can reiterate in this paper is that \vDome\ is a
poor estimator of the reference flat: its use would generate 10's
of mmag of error into DECam photometry.  Two reasons are
well-known: the illumination spectrum of \vDome\ never matches a
desirable choice for $F_{\rm ref}$; and ``flat''-field sources usually
are not quite uniform in surface brightness.  There are two other
issues that become more important for wide-field imagers: first,
\vDome\ contains a factor of \vOmega, which should not be present,
according to \eqq{rflat}.  Second, our reference flat should only
include light that has been properly focussed by the telescope onto
the array, since these photons are the only ones that are counted in
photometric measures.  But many photons reach the focal plane through
stray reflections and scattering---several percent in the case of
DECam.  The photons that arrive at the detector out of
focus\footnote{More exactly, outside of a
nominal 6\arcsec\ aperture.} will be generically referred to as
``stray light,'' and their signal cannot be distinguished from
focussed light in the image of a uniform screen.  

To remedy the problems with \vDome, we produce a \emph{star flat}
image $\vSFlat(t)$ such that 
\begin{equation}
\vrr(t) = \vDome \times \vSFlat(t).
\end{equation}
The star flat has the task of removing from \vDome\ those features
that are not reflective of true response to focussed stellar photons.
Note that we have no time dependence on \vDome.
We find the instrument response to
be far more stable from night to night than the dome illumination
system, a tribute to the design and implementation of the camera \citep{estrada}.
Thus the use of daily dome exposures to flatten nightly images
actually increases instability of the photometric calibration.  We instead
produce a single \vDome\ per filter to apply to an entire season's images.
Indeed for this paper, where we are investigating multi-year trends in
instrument response, we use a single \vDome\ for the entire history of
the instrument.

Why bother with \vDome\ at all?  We will be solving for \vSFlat\ by
optimizing the parameters of some functional form for it.  There are
some small-scale features in $\vrr(t)$ that are not easily
parameterized, such as spots and scratches, that \vDome\ will
capture.  The cosmetic quality of DECam CCDs is, however, very high,
and we might in fact be better off eliminating \vDome.  Daily dome
flats remain useful, however, for identifying transitory phenomena
such as dust or insects on the optics.

\subsection{Constraining the star flats and color terms}
We take the philosophy that the best way to characterize the
photometric behavior of DECam is to examine real on-sky stellar
photometry. We 
derive the best form for the star flats $\vSFlat(t)$ and the color
terms $\vc(t)$, by these are functions of some parameters \vpi, 
and then finding the
parameters which minimize the sum
\begin{equation}
\chi^2 = \chi^2_\star + \chi^2_{\rm atm}
\label{chisq1}
\end{equation}
for a set of stellar photometric observations with the camera.
The first term quantifies the internal consistency of the
stellar magnitudes and the second term the adherence of the solution
to the atmospheric prior.  To enforce internal consistency we
begin by rearranging \eqq{mag1}
with the following definitions:
\begin{align}
m_\alpha & = m_{\alpha t} + \Delta m_{\alpha t}(\vpi) \\
\label{minst}
m_{\alpha t} & \equiv m_1 - 2.5\log_{10} \sum_{ {x \in \alpha}}
\frac{\vRate_\star(t)}{\vDome} \\
\nonumber
\Delta m_{\alpha t}(\vpi) = \Delta m(\vx_{\alpha t}, p_\alpha, t; \vpi) & = 
2.5\log_{10} \vSFlat(t) + p_\alpha \vc(t)
                                              + Ap(t)  \\
         & \equiv S(\vx_{\alpha t};\vpi_S) + p_\alpha \, c(\vx_{\alpha
            t}; \vpi_c) + R(\vx_{\alpha t}, p_\alpha;\vpi_R).
\label{deltam}
\end{align}
The instrumental magnitude $m_{\alpha t}$ is assigned in \eqq{minst}
to the
aperture sum for star $\alpha$ in image $t$ flattened using only the
\vDome\ portion of the reference flat (with an arbitrary zeropoint $m_1$).
In \eqq{deltam}, we rearrange the total magnitude correction $\Delta
m$ being applied to the instrumental magnitude into three terms:
the \emph{instrument} portion $S$ which is constant in time, and
independent of color; the \emph{color} term $c$, which we now also
assume is independent of time; and an
\emph{exposure}
term $R$, which is (nearly) constant across the field
of view and absorbs the temporal variation of the reference flat.
Each of these is now written as a scalar function in magnitude units, explicitly
dependent on the array position $\vx_{\alpha t}$ at which the star is
observed. The total parameter set $\vpi$ of the response model is the
union of the $\vpi_S, \vpi_c,$ and $\vpi_R$ of the three terms.
This parameter split follows the approach of
{\sc scamp} software \citep{scamp}.

The measure of internal consistency is now
\begin{equation}
\label{chistar}
\chi^2_\star \equiv \sum_{\alpha} \sum_{t  \in \alpha}
\frac{\left[ m_{\alpha t} + \Delta m_{\alpha t}(\vpi) 
    - m_\alpha\right]^2}{\sigma^2_{\alpha t} + \sigma^2_{\rm sys}}.
\end{equation}
The photometric uncertainty $\sigma_{\alpha t}$ is calculated from the
detector noise and photon noise at the stellar aperture.  The
$\sigma_{\rm sys}$ places a floor on the uncertainty to cap to weight
placed on any single stellar detection. Its use means that we should
not expect $\chi^2$ to follow a true $\chi^2$ distribution, so we will
evaluate the quality of the model by other means. In practice we will
set $\sigma_{\rm sys}=2$~mmag, which we will find in
\S\ref{errorstats} is the excess of the measured photometric variance
over that expected from shot noise and read noise.
The $\chi^2$
sum is over all observations of each star, with a free parameter
$m_\alpha$ for the true magnitude of each star.  

The form of $\chi^2_{\rm atm}$ depends on the model adopted for
atmospheric extinction.  \citet{fgcm} make use of a physical model
parameterized by the concentrations of various atmospheric constituents, but in
this work we will adapt the simple traditional approach of assigning a
zeropoint $k_{n0},$ an airmass coefficient $k_{n1}$, and a color term
$k_{n2}$ to each night $n$ of data.  We will find it beneficial to
include a fourth term with coefficient $k_{n3}$ multiplying some
variable $A_t$ thought to be a proxy for variability in the
aperture correction $Ap(t)$.  The atmospheric prior is thus
quantifying our expectation that the magnitude correction $\Delta
m(\vx_s, p_s, t)$ for sample point $s$ at some fixed focal-plane location $\vx_s$ and color(s)
$p_s$ in exposure $t$ taken on night $n$ should obey
\begin{equation}
\Delta m(\vx_s, p_s, t) = k_{n0} + k_{n1} (X_t-1) + k_{n2} \cdot p_s +
k_{n3} \cdot A_t.
\label{atmo1}
\end{equation}
Here $X_t$ is the mean airmass on the line of sight for exposure $t$.
We will investigate in Section~\ref{atmosphere} the efficacy of some choices
for aperture-correction proxy $A_t$. 

Next we assume that all exposures $t\in n$ on a given night $n$ should obey
this equation to an RMS accuracy $\sigma_n,$ a measure of the accuracy
of our model and stability of the atmosphere during the time span of
the ``night'' (which need not be exactly one night).
We measure adherence to this prior expectation by sampling each
exposure's solution with pseudo-stars of two different colors:
\begin{equation}
\chi^2_{\rm atm} = \sum_n \sum_{t \in n} \sum_{p_s=0,1} 
\frac{ \left[ \Delta m(\vx_s, p_s, t) - k_{n0} - k_{n1} (X_t-1) -
    k_{n2} \cdot p_s -
k_{n3} \cdot A_t \right]^2}{\sigma^2_n}.
\label{atmprior}
\end{equation}
The values $k_{nj}$ form additional free parameters of the response
model.  Like all the parameters of our model, they can be held fixed
to \textit{a priori} values if desired.

The grand scheme is that we will obtain, on clear nights, exposures of
rich but uncrowded stellar fields, and extract high-$S/N$ aperture
photometry to establish $m_{\alpha t}$ for a large number of stars.
If each individual star is measured at a wide span of locations
$\vx_{\alpha t}$ on the focal plane, and they span a range of colors
$p_\alpha$, then we can solve simultaneously the parameters of $S$ and
$c$.  Time variation $R$ can be allowed as well, to probe
instabilities over the span of the observations.  The atmospheric
prior will constrain these time variations and break some degeneracies
that we discuss below.

\subsection{Terminology}
\label{terminology}
The \texttt{C++} program \photofit\ executes the $\chi^2$ minimization
defined in the previous section.  The code for this and related
programs is available at \url{https://github.com/gbernstein/gbdes},
where one can find documentation of its installation and use.  Here we
provide an overview of the concepts and algorithms implemented in the
code.   \photofit\ shares many characteristics and code sections with
the \wcsfit\ program that produces the DECam astrometric model.
Both codes are generically applicable to array cameras, not just
DECam.  \citet{decamast} provide a description of \wcsfit, so in this
paper we will be concise in our descriptions of aspects of \photofit\ that
are shared with \wcsfit.

Before proceeding further, we define the \photofit\ terminology:
\begin{itemize}
\item A {\bf detection} is a single measurement of a stellar magnitude (flux),
  $m_{\alpha t}$ in the notation above.  It has an associated
  measurement noise $\sigma_{\alpha t}.$
\item A {\bf device} is a region of the focal plane over which we
  expect the photometric calibration function $\Delta m$ be continuous,
 \ie\ one of the CCDs
  in the DECam focal plane.  Every detection belongs to exactly one device.
\item An {\bf exposure} comprises all the detections obtained
  simultaneously during one opening of the shutter.  The exposure
  number is essentially our discrete time variable $t$.
\item An {\bf extension} comprises the detections made on a single
  device in a single exposure.  
\item A {\bf catalog} is the collection of all detections from a
  single exposure, \ie\ the union of the extensions from all the
  devices in use for that exposure.
\item A {\bf band} labels the filter used in the observation.  Every
  exposure has exactly one band.  All detections of a given
  non-variable star in the
  same band should yield the same magnitude $m_\alpha$.
\item An {\bf epoch} labels a range of dates over which the physical
  configuration of the instrument, aside from filter choice and the
  pointing of the telescope, is considered (photometrically)
  invariant.  Every exposure belongs to exactly one epoch.
\item An {\bf instrument} is a given configuration of the telescope
  and camera for which we expect
  the instrumental optics to yield an invariant photometric solution.
  In our analyses an instrument is specified by a combination of band
  and epoch.  Every exposure is associated with exactly one instrument.
  This is the same definition as used in \scamp.
\item A {\bf match}, sometimes called an {\bf object}, comprises all
  the detections that correspond to a common celestial source.  We will only
  make use of stellar sources, since aperture corrections are
  ill-defined for galaxies. \photofit\ only matches detections in the
  same band, since we require matched detections to have common true
  $m_\alpha.$ 
\item A {\bf field} is a region of the sky holding the detections from
  a collection of exposures.  Every exposure is associated with exactly
  one field.  A detection is only matched to other detections in the
  same field.
\item A {\bf photomap} is a magnitude transformation $m_{\rm
    in}\rightarrow m_{\rm out}$.  All of our photomaps take the form
  $m_{\rm out} = m_{\rm in} + \Delta m(\vx_{\rm dev}, \vx_{\rm fp},
  p; \vpi).$ The magnitude shift can depend on the position of the
  star on the device (in pixels), or on the position $\vx_{\rm fp}$ of the device in
  a coordinate system centered on the optic axis and continuous across
  the focal plane.  If there is dependence
  on the object color $p$, the photomap is called ``chromatic.''  The
  photomap can also have controlling parameters \vpi.  The
  transformation from instrumental magnitude $m_{\alpha t}$ to
  calibrated magnitude $m_\alpha$ 
  is realized by compounding several photomaps, and a compounded map
  is also called a photomap.  \photofit\ models define one photomap
  per extension, but these maps may be have components that are common
  to multiple extensions.
\end{itemize}

\subsection{Available maps}
\label{maps}
\photofit\ allows each extension's photomap to be composed of a series
of constituent ``atomic'' maps.  Given that all maps in use correspond
to additive magnitude shifts $\Delta m$, this amounts to simply
summing the values of $\Delta m$ from each component.

\photofit\ follows the definitions in Section~\ref{terminology} by making
each magnitude transformation or element thereof an
instance of an abstract \texttt{C++} base class \texttt{PhotoMap}.
Each has a \texttt{type}, a unique \texttt{name} string, and has a
number $\ge0$ of free parameters controlling its
actions.  \texttt{PhotoMap} instances can be (de-)serialized (from) to
ASCII files in \textsc{YAML} format, easily read or written by
humans.  The \photofit\ user can compactly specify the functional form
desired for all of the observations' photomaps, either supplying
starting parameter values or taking defaults.
The primary output of \photofit\ is another \textsc{YAML} file
specifying all of the photomaps and their best-fit parameters.

These are the implementations of \texttt{PhotoMaps}:
\begin{itemize}
\item The \texttt{Identity} map, $\Delta m=0,$ has no free parameters.
\item \texttt{Constant} maps have the single free parameter $\Delta m = \Delta m_0.$
\item \texttt{Polynomial} maps have $\Delta m$ as a polynomial
  function of the detection position.  One can select either device
  (pixel) coordinates or focal-plane coordinates to be the arguments.
  The user also specifies either distinct maximum orders for the $x$
  and $y$ in each term of the polynomial, or the maximum sum of the
  $x$ and $y$ orders.  The polynomial coefficients are the parameters.
\item \texttt{Template} maps apply magnitude shifts based on lookup
  tables that are functions of either $x, y,$ or radius from
  some pre-selected center $\vx_c$:
\begin{align}
  \Delta m & = s \cdot f(x), \\
  \Delta m & = s \cdot f(y), \text{or} \\
  \Delta m & = s \cdot f\left(|\vx-\vx_c| \right).
\end{align}
There is a single free parameter, the scaling 
$s$.  The template function $f$ is defined as linear interpolation
between values $v_j$ at nodes $a_0 + j\,\Delta a$ for $0\le j \le N$.
The user again selects whether device or focal-plane coordinates
should be used.
\item \texttt{Piecewise} maps are functionally identical to the
  \texttt{Template} map, except that the nodal values $v_j$ are the
  free parameters, and the scaling is fixed to $s=1.$  
\item A \texttt{Color} term is defined by 
\begin{equation}
\Delta m = \left(p-p_{\rm  ref}\right)\times\Delta m_{\rm targ}(\vx_{\rm
  dev}, \vx_{\rm fp}; \vpi),
\end{equation}
where $p_{\rm ref}$ is a reference color and the target photomap
$\Delta m_{\rm targ}$ is an instance of any non-chromatic atomic photomap.
The parameters of the \texttt{Color} map are those of its target.
\item \texttt{Composite} maps realize composition
  of a specified
  sequence of any of photomaps (including other \texttt{Composite} maps).
  The parameters of the composite are the concatenation of those of
  the component maps.
\end{itemize}

\subsection{Degeneracies}
\label{degeneracies}
When minimizing $\chi^2$ we must be aware of degeneracies whereby
\vpi\ can change while $\chi^2$ is invariant.  Such degeneracies will
lead to (near-)zero singular values in the normal matrix \matA\ used
in the solution for \vpi\ (Section~\ref{algorithms}), and failures or
inaccuracies in its inversion.  

For astrometric solutions described in \citet{decamast}, many
degeneracies 
are resolved using
an external reference catalog with absolute sky positions for a selection
of stars.  In general no such reference catalog will be available for
magnitudes in our camera's natural bandpasses.  
Thus the photometric case becomes less straightforward, and indeed it is these
degeneracies in the internal-consistency constraints that require use
of an atmospheric prior for an unambiguous solution.
We will assume in this discussion that the
photometric model $\Delta m=S+R_t$ for each extension is a
device-based instrumental function $S$, plus an exposure-specific function
$R_t,$ both potentially being functions of focal-plane coordinates.

\subsubsection{Absolute calibration}
The simplest degeneracy is a shift in all stellar magnitudes.
$\Delta m\rightarrow \Delta m + \delta m.$  A corresponding shift
$m_\alpha\rightarrow m_\alpha + \delta m$ leaves $\chi^2_\star$
invariant, and $k_{n0}\rightarrow k_{n0}+\delta m$ leaves $\chi^2_{\rm
  atm}$ invariant.  This is simply our ignorance of the absolute
photometric calibration.  The absolute calibration can be constrained
by adding to the atmospheric prior in \eqq{atmprior} a fictitious 
``night'' for which we hold all the nightly parameters $k_{nj}$ fixed.
We place a single exposure into this ``absolute prior,'' which
essentially anoints this exposure as a reference field.  The
absolute-calibration degeneracy is then also broken
for any exposures matching stars with the reference exposure, or having
been observed in the same night as the reference.
We may require more than one exposure in the absolute prior if we have
observed two disjoint fields but never on the same night.  
We need distinct absolute priors for each band.
A script provided with \photofit\ can deduce a workable set of absolute
priors for each band given the fields and dates of all exposures.

\subsubsection{Color shift}
A color-dependent shift $\Delta m\rightarrow \Delta m + pC$ for some constant
$C$ is also undetectable in $\chi^2$, since we can counter with
$m_\alpha \rightarrow m_\alpha + p_\alpha C.$ 
We again need to select one or more exposures to serve as absolute
references for the color scale.  This is easily accomplished by adding
sample points of two different colors to the absolute prior.  Indeed
there are color-term analogs to all of the degeneracies discussed
here, and each requires incorporation of some chromatic version of the
degeneracy-breaking constraint.  \photofit\ does not currently
identify all of these automatically, so the user must take care that
his/her model does not introduce color-scale degeneracies.

\subsubsection{Gradient}
Consider a set of exposures with pointings at $\vx_t$ such that 
stars at sky coordinates $\vx_\alpha$ appear
at focal-plane coordinates $\vx_{\alpha t}=\vx_\alpha-\vx_t$ on each
exposure.  The following transformations of our instrument solution,
exposure solution, and source magnitudes, respectively, leave $\chi^2_\star$
unchanged for any linear gradient ${\bf g}$:
\begin{align}
S(\vx_{\rm fp}) & \rightarrow S(\vx_{\rm fp}) + {\bf g}\cdot \vx_{\rm
                  fp}, \\
R_t & \rightarrow R_t + {\bf g}\cdot \vx_t \\
m_\alpha & \rightarrow m_\alpha + {\bf g} \cdot \vx_\alpha.
\end{align}
In the flat-sky limit this degeneracy is exact, and the
internal-consistency constraint is seen to be completely insensitive
to gradients across the focal plane in the instrumental solution
$S(\vx_{\rm fp})$.  The atmospheric prior suppresses this degeneracy,
however, because it penalizes the gradient in $R_t$ across the sky
that is needed for degeneracy in $\chi^2_\star.$
If we sample each exposure solution at $\vx_s=0$, then the
sum in \eqq{atmprior} becomes, on an otherwise perfectly-behaved
night,
\begin{equation}
\chi^2_{\rm atm} = \sum_t \left[\frac{{\bf g} \cdot \left(\vx_t-\bar\vx_t\right)}{\sigma_n}\right]^2,
\end{equation}
where $\bar\vx_t$ is the mean sky position of the night's pointing.
The gradient should hence be suppressed to a level roughly given by
$\sigma_n$, the typical fluctuation in atmospheric transmission,
divided by the angle spanned by observations on a given night.  This
shows why suppression of large-scale modes in the photometric
solutions requires one to make occasional large-angle slews during
nights with a stable atmosphere---it defeats the gradient degeneracy
and also helps one separate spatial and temporal changes in the
atmosphere.  In
Section~\ref{atmosphere} we will characterize $\sigma_n$ empirically.

There is a higher-order version of the gradient degeneracy if the
exposure solution has polynomial freedom at order $q>1$ and the
exposure solution has freedom at order $q-1$. Such degeneracies are
suppressed by a proper atmospheric prior, usually more strongly than
the simple gradient.  We also note that the gradient degeneracy is
mathematically broken by the curvature of the sky, moreso if the survey
encircles the sky--- but large-scale modes remain weakly constrained.

\subsubsection{Exposure/instrument trades}
Any time the functional forms of $S$ and $R$ admit a transformation of
the form
\begin{align}
S & \rightarrow S + f(\vx_{\rm fp}) \\
R_t & \rightarrow R_t - f(\vx_{\rm fp}) 
\end{align}
for some function $f$, this leaves both $\chi^2_\star$ and
$\chi^2_{\rm atm}$ invariant. The \photofit\ code
searches for cases where multiple \texttt{Constant} or
\texttt{Polynomial} atomic map elements are composited into any
exposures' photomaps and are hence able to trade their terms.
This degeneracy can be broken by setting one of the exposure maps
$R_t$ to the \texttt{Identity} map.  \photofit\ will do this
automatically if the user's configuration leaves such degeneracies in
place.  Note that this means the ``instrumental'' solution $S$
incorporates one exposure's manifestation of any time-varying
response.  This is just a semantic issue, it does not affect the resultant
photometric solutions.

\subsubsection{Unconstrained parameters}
\photofit\ checks the normal matrix \matA\ (Section~\ref{algorithms}) 
for null rows that arise
when a parameter does not act on any observations.  In this case the
diagonal element on this row is set to unity, which stabilizes the
matrix inversion and freezes this (irrelevant) parameter in further
iterations.
When there are a finite but insufficient number of observations to
constrain the model, \photofit\ will fail in attempting a Cholesky decomposition of a
non-positive-definite  \matA.  In this case \photofit\ provides the
user with a description of the degenerate eigenvector(s), then quits.
The user must identify the problem, which usually can be traced back
to inclusion of defective exposures and/or failure to resolve one of
the degeneracies above.

\subsection{Algorithms}
\label{algorithms}

\photofit\ optimizes the model parameters \vpi\ with simple algorithms
once the model is specified and the data are read. Each
\texttt{PhotoMap} implementation is capable of calculating $\partial
\Delta m / \partial \vpi$ and we approximate $\chi^2$ with the
usual quadratic form
\begin{align}
\chi^2 & \approx \chi^2(\vpi_0) + 2{\bf b} \cdot \Delta\vpi +
         \Delta\vpi \cdot \matA \cdot \Delta \vpi, \\
 b_\mu & \equiv \frac{1}{2} \frac{\partial \chi^2}{\partial \pi_\mu} =
         \sum_{\alpha, t}
         w_{\alpha t}\left[m_{\alpha t}+\Delta m_{\alpha t}(\vpi_0) - \bar m_\alpha \right]
  \cdot \left(\frac{\partial \Delta m_{\alpha t}}{\partial \pi_\mu}
   - \frac{\partial \bar m_\alpha}{\partial \pi_\mu} \right)\\
A_{\mu\nu} & \equiv \sum_{\alpha t} w_{\alpha t}\left[\left(\frac{\partial \Delta m_{\alpha t}}{\partial \pi_\mu}
   - \frac{\partial \bar m_\alpha}{\partial \pi_\mu} \right)
             \cdot 
\left(\frac{\partial \Delta m_{\alpha t}}{\partial \pi_\nu}
   - \frac{\partial \bar m_\alpha}{\partial \pi_\nu}
             \right)\right]. \\
w_{\alpha t} & \equiv \left( \sigma_{\alpha t}^2 + \sigma_{\rm
               sys}^2\right)^{-2} \\
\bar m_\alpha & \equiv \frac{ \sum_{t \in \alpha} w_{\alpha t} \left(
                m_{\alpha t} + \Delta m_{\alpha t}\right)}{\sum_{t \in
                \alpha} w_{\alpha t}}.
\end{align}
\photofit\ does not treat the true magnitudes $m_\alpha$ 
as free parameters.  Instead the dependence of the mean of the
measurements $\bar m_\alpha$ upon the parameters is propagated
directly into the normal equation, in practice executing an analytic
marginalization over $m_\alpha$.

The calculation of ${\bf b}$ and \matA\ is the most computationally
intensive part of \photofit.  The summation for matches is distributed
across cores using \textsc{OpenMP} calls.  Updates to \matA\ are
sparse, though the final matrix is dense. 
\photofit\ first attempts the Newton iteration
\begin{equation}
\label{Newton}
\vpi \rightarrow \vpi - \matA^{-1} {\bf b}.
\end{equation}
The solution is executed using a multithreaded Cholesky
decomposition after preconditioning \matA\ to have unit diagonal
elements. 

The Newton step is iterated until $\chi^2$ no longer decreases by more
than a chosen fraction.  Should $\chi^2$ increase during an iteration,
or fail to converge within a selected number of steps, then the
minimization process is re-started using a Levenberg-Marquart
algorithm based on the implementation by \citet{recipes}.

\subsubsection{Outlier rejection}
The \photofit\ solutions must be robust to fluxes
contaminated by unrecognized cosmic rays or defects, and to stars
with variable magnitude over the observation timespan.
Outlier rejection is done using standard $\sigma$-clipping
algorithms. A clipping threshold $t$ is specified at input.  After
each $\chi^2$ minimization, a rejection threshold is set at $t
\sqrt{\chi^2/\textrm{DOF}}$.
Detections whose residual to the fit (in
units of $\sigma_{\alpha t}$) exceeds the threshold are discarded.  At most one
outlier per match is discarded at each clipping iteration.
Outlier clipping is alternated with $\chi^2$ minimization until the
clipping step no longer reduces the $\chi^2$ per degree of freedom by
a significant amount.

Outlier exposures are clipped from consideration in $\chi^2_{\rm
  atm}$ by a similar process, in case there are exposures with
erroneous shutter times, occultation by the telescope dome, or other
freak occurrences that perturb the exposure zeropoint.

\subsubsection{Procedure}
The steps in the photometric solution process are as follows:
\begin{enumerate}
\item A preparatory Python program 
reads an input YAML configuration file specifying the desired input
catalog files, 
plus the definitions of the fields, epochs, and instruments.  It then
collects from all the catalogs and their headers any information
necessary to construct tables of extensions, 
devices, exposures, and instruments.  This includes extracting a
serialized world coordinate system, usually as produced by \scamp\ and stored in
the headers of the FITS catalog extensions.  
\item A second preparatory program reads all the detections from the
  input catalogs, applying any desired cuts for $S/N$ and stellarity,
  and then runs
a standard friends-of-friends algorithm to identify all detections
with matching sky positions.
The id's of all
groups of matching detections are then stored in another FITS table.
\item The user or a preparatory script creates another YAML file that
  specifies the exposures to be assigned to each night of the
  atmospheric prior, which exposures to place in a pseudo-night to
  constrain the absolute magnitude and color scales.  The initial
  values of the nightly parameters $k_{nj}$ are also specified in this
  file, and the file also specifies which $k_{nj}$ are free to vary
  during optimization.
\item \photofit\ starts by ingesting the input FITS tables and creating
  the structures defining instruments, devices, exposures, and
  extensions.
\item The YAML file specifying the photomaps to be applied to each
  extension is parsed, and a \texttt{PhotoMap} is created with
  specified or defaulted parameters.  Any of the map elements may have
  its parameters frozen by the user, the remainder are the free
  parameters of our model.
\item \photofit\ checks the map configuration for 
  exposure/instrument degeneracies, attempting to break any
  by setting one or more exposures' maps to  \texttt{Identity}. 
\item The $m_{\alpha t}$ and $\sigma_{\alpha t}$ of all detections that are part
  of useful matches are extracted from their source catalogs.  For any
  detections whose maps include color terms, we require a measurement
  from a color catalog to be matched to the same object.  The color
  catalog is read at this point.
\item A requested fraction of the matches are excluded from the fit at
  random.  These reserved matches can be used later to validate the fit.
\item Any exposures containing insufficient detections are removed
  from the fit.
\item The atmospheric prior configuration is read from a YAML file and
  the free parameters of each night's model added to \vpi.
\item The iteration between $\chi^2$ minimization and
  $\sigma$-clipping begins.  At each iteration, \matA\ is checked for
  null rows as noted in Section~\ref{degeneracies}, which are altered
  so as to freeze the associated parameter.  If \matA\ is not
  positive-definite, \photofit\ reports the nature of the associated
  degenerate parameters, then exits.
\item As the iterations near convergence, further iterations are
  allowed to clip up to one outlier exposure per night in the
  atmospheric prior.
\item The best-fit photometric model is written to an output YAML file.
\item After completion of the fit, the best-fit map is applied to both
  the fit and reserved matches.  The $\sigma$-clipping algorithm is
  applied iteratively to the reserved matches.
\item The RMS residual and $\chi^2$ statistics are reported for the
  un-clipped detections on each exposure.
\item The input, output, and best-fit residual for every detection are
  written to an output FITS table for further offline analyses.
  Another output file reports the parameters of the nightly
  atmospheric models and the residuals of each exposure to them.
\end{enumerate}
In typical usage the entire process would be first executed with an
achromatic model for $g$ and $i$ bands.  The $g-i$ colors of the
matched stars can then be determined, 
and the \photofit\ steps re-run with a chromatic model.

\section{DECam data and model}
\label{model}
We now proceed to apply the methodology and code in the previous
section to derivation of star flat functions $S(\vx)$ and color terms
$c(\vx)$ for DECam.  
\subsection{Observations}
\label{data}
Our primary constraints on the response function come from specialized
``star flat'' observing sequences, 
typically obtained during engineering nights in bright-moon
conditions. A typical star flat session consists of $22\times30$~s
exposures per filter in a field at modest Galactic latitude, where bright stars are
abundant but sparse enough for successful aperture photometry.  The
exposure pointings are dithered by angles from 10\arcsec\ up to the
1\arcdeg\ radius of the FOV.  It is essential for the success of
internal calibration to have a wide variety of displacement vectors
on the focal plane between the detections of a given star.  With
25--30~s between exposures for readout and repointing, the total clock
time for a star flat sequence is about 20--25~minutes per filter, or
2~hours to complete the $grizY$ filter set used by DES.  A star
flat session yields $\approx 5\times10^5$ matched detections per filter at
$1000\gtrsim S/N>30.$

Star flat sequences have been taken once per few months since
commissioning of DECam in October 2012.  Table~\ref{epochs} lists the dates
and conditions of the star flat sequences used in this analysis,
namely those to date free of clouds and instrument problems.

\begin{deluxetable}{lccc}
\tabletypesize{\footnotesize}
\tablewidth{0pt}
\tablecolumns{8}
\tablecaption{Star flat observing sequences through Feb 2017\label{epochs}}
\tablehead{
\colhead{Epoch\tablenotemark{a}} & 
\colhead{Field} &
\colhead{$D_{50}$\tablenotemark{b}} &
\colhead{Airmass}
}
\startdata
\texttt{20121120}\tablenotemark{c} &  0640--3400 & 2\farcs09 & 1.04 \\
\texttt{20121223} &   0730--5000 & 1\farcs04 & 1.06 \\
\texttt{20130221} &   1327--4845 & 1\farcs12 & 1.06 \\
\texttt{20130829} &   1900--5000 & 1\farcs10 & 1.07 \\
\texttt{20131115} &   0640--3400& 1\farcs41 & 1.09 \\[3pt]
\textit{2013 Nov 30} & \multicolumn{3}{c}{\textit{CCD S30 fails}\tablenotemark{e}} \\[3pt]
\texttt{20140118} &   1327--4845 & 1\farcs33 & 1.33 \\
\texttt{20140807} &   1327--4845 & 1\farcs43 & 1.32 \\
\texttt{20141105}\tablenotemark{d} &   0640--3400 & 1\farcs28 & 1.01 \\
\texttt{20150204} &   1327--4845 & 0\farcs88 & 1.31 \\
\texttt{20150926} &   2040--3500 & 1\farcs19 & 1.01 \\
\texttt{20160209} &   0730--5000 & 1\farcs25 & 1.07 \\
\texttt{20160223} &   1327--4845 & 1\farcs10 & 1.24 \\
\texttt{20160816} &   1900--5000 & 1\farcs08 & 1.06 \\
\texttt{20161117} &   0640--3400 & 1\farcs00 & 1.20 \\[3pt]
\textit{2016 Dec 28} & \multicolumn{3}{c}{\textit{CCD S30 revives}\tablenotemark{e}} \\[3pt]
\texttt{20170111} &   0640--3400 & 1\farcs32 & 1.22 \\
\texttt{20170214} &   1327--4845 & 1\farcs11 & 1.06 
\enddata
\tablenotetext{a}{The local date at start of the night when the star flat
  exposures were taken or event occurred.}
\tablenotetext{b}{Median half-light diameter of the point spread
  function for the $i$-band exposures in the sequence.}
\tablenotetext{c}{$zY$ star flats were taken on the following night.}
\tablenotetext{d}{$zY$ star flats were taken on 10 Nov.}
\tablenotetext{e}{The number of 
  functional CDDs on DECam dropped from 61 to 60 with the failure of
  S30 on the indicated date. The dead CCD came back to life 3 years
  later, as noted.
  Plots in this paper hence vary in the number of CCDs in use. S30 is
  at top dead center of the focal plane
  images in this paper.}
\label{starflats}
\end{deluxetable}

\subsection{Stellar flux measurement}
Raw images from each star flat exposure are
run through the DES detrending steps described in \citet{detrend} up
to the point where the star flat correction would be applied:
linearization of images, crosstalk removal, conversion from ADU to
photocarrier counts, correction for
the ``brighter-fatter effect'' \citep{gruen}, debiasing, and
division by dome flats, and subtraction of sky and fringe signals.
Sources are detected and measured using \textsc{SExtractor}
\citep{sextractor}.  For the following analyses we filter the catalogs
for sources with no \textsc{SExtractor} flags set, no defective,
saturated, or cosmic-ray-flagged pixels
within the isophote, with
\texttt{MAGERR\_AUTO}$<0.03,$ indicating signal-to-noise ratio
$S/N\gtrsim30,$ and with $|\texttt{SPREAD\_MODEL}|<0.003$ to select
only stellar sources.  The flag cut removes objects that overlap
detected neighbors.  

Any photometric calibration is tied to the particular algorithm for
extracting stellar fluxes from the image.  Optimal $S/N$ of the
extracted flux comes from PSF-fitting photometry. But ideally we
would also like to capture a ``total'' magnitude for the star,  counting
all the photons that make it to our detector.  In this way our flux measures
the efficiency of the detector and atmosphere with minimal dependence
on the PSF, its variations across the FOV, or
our ability to measure the PSF.
We opt to favor PSF robustness over $S/N$ by counting flux in a
circular aperture of diameter 6\arcsec,
generously larger than the typical $\approx1\farcs0$ FWHM of the
seeing.  The PSF keeps going past 6\arcsec, all the way to
the horizon.  In Section~\ref{atmosphere} we infer that temporal
variation in camera response within a cloudless night is usually
dominated by fluctuations in the amount of light scattered outside our
nominal aperture, i.e. the $Ap(t)$ term in \eqq{mag1}.  While we
cannot integrate the PSF to the horizon, we investigate whether the
amount of light just outside our nominal aperture is informative of
fluctuations in the total aperture correction.
We will therefore use a second aperture magnitude from \sextractor\ 
to define for exposure $t$
\begin{equation}
A_t = {\rm Median}\left[ {\tt MAG\_APER}(8\arcsec) -
  {\tt MAG\_APER}(6\arcsec)\right].
\label{apcorrdef}
\end{equation}
The median is over bright stars, yielding
a measure of $A_t$ that is high-$S/N$ and robust to
neighboring objects even as the outer aperture grows.

Note that it is perfectly feasible for the photometric calibration to
be defined via stellar aperture fluxes while science measures use
lower-noise PSF-fitting fluxes: the latter can be normalized to the
former once a PSF model is in hand.

Background estimation is also critical for high-precision photometry.
We set
\texttt{BACKPHOTO\_TYPE$=$LOCAL} in \textsc{SExtractor} so that the sky
level is set to the mode in a rectangular ``annulus'' of thickness $\approx
6\arcsec$ surrounding the stellar isophotes.
Aperture fluxes are converted to magnitudes $m_{\alpha t}$ as per
\eqq{minst}, and \textsc{SExtractor} also provides an estimate of the
shot-noise uncertainty $\sigma_{\alpha t}$ on this magnitude.  Note that
\textsc{SExtractor} does not propagate uncertainties in sky determination
into $\sigma_{\alpha t}.$

\subsection{The DECam photometric model}
Table~\ref{mapelements} lists the components that we find necessary
for a photometric model of the DECam \vSFlat\ to
approach mmag repeatability in DECam photometry in the star flat
sequences. We first examine the static ``instrumental'' components of
the model, and will attend to the time-variable components in later
sections. 

\begin{deluxetable}{lccc}
\tablewidth{0pt}
\tablecolumns{8}
\tablecaption{Components of the DECam photometric model}
\tablehead{
\colhead{Description} &
\colhead{Name} &
\colhead{Type} &
\colhead{Max.\ size (mmag)}
}
\startdata
Tree ring distortion &
$\langle\textit{band}\rangle\texttt{/}\langle\textit{device}\rangle\texttt{/rings}$
& \texttt{Template} (radial)  &$\pm20$ \\
Serial edge distortion &
$\langle\textit{band}\rangle\texttt{/}\langle\textit{device}\rangle\texttt{/lowedge}$
& \texttt{Template} (X) & $\pm10$ \\
Serial edge distortion &
$\langle\textit{band}\rangle\texttt{/}\langle\textit{device}\rangle\texttt{/highedge}$
& \texttt{Template} (X) & $\pm10$ \\
Optics/CCDs &
$\langle\textit{band}\rangle\texttt{/}\langle\textit{device}\rangle\texttt{/poly}$
& \texttt{Polynomial} (order$=3$) & $\pm40$ \\
Color term &
$\langle\textit{band}\rangle\texttt{/}\langle\textit{device}\rangle\texttt{/color}$
& \texttt{Color}$\times$\texttt{Linear} & $\pm5$ (25 in $g$) \tablenotemark{a}  \\
Exposure zeropoint &
$\langle\textit{exposure}\rangle$
& \texttt{Constant} & (large) \\
Exposure color &
$\langle\textit{exposure}\rangle\texttt{/color}$
& \texttt{Color}$\times$\texttt{Constant} & $\pm4$ \tablenotemark{a} \\
Extinction gradient &
$\langle\textit{exposure}\rangle\texttt{/gradient}$
& \texttt{Polynomial} (order$=1$) & $\pm20$ ($g,\,X=2$) \\
Long-term drift &
$\langle\textit{epoch}\rangle\texttt{/drift}$
& \texttt{Polynomial} (order$=4$) & $\pm6$ \\
CCD QE shift\tablenotemark{b} & 
$\langle\textit{epoch}\rangle\texttt{/ccd}$
& \texttt{Constant}  & $\pm 5$
\enddata
\tablenotetext{a}{The color terms are in units of mmag per magnitude
  of $g-i$ color.}
\tablenotetext{b}{Only for $Y$ band.}
\label{mapelements}
\end{deluxetable}

\subsubsection{Tree rings and glowing edges}
\begin{figure}[!h]
\begin{center}
\includegraphics[height=4in]{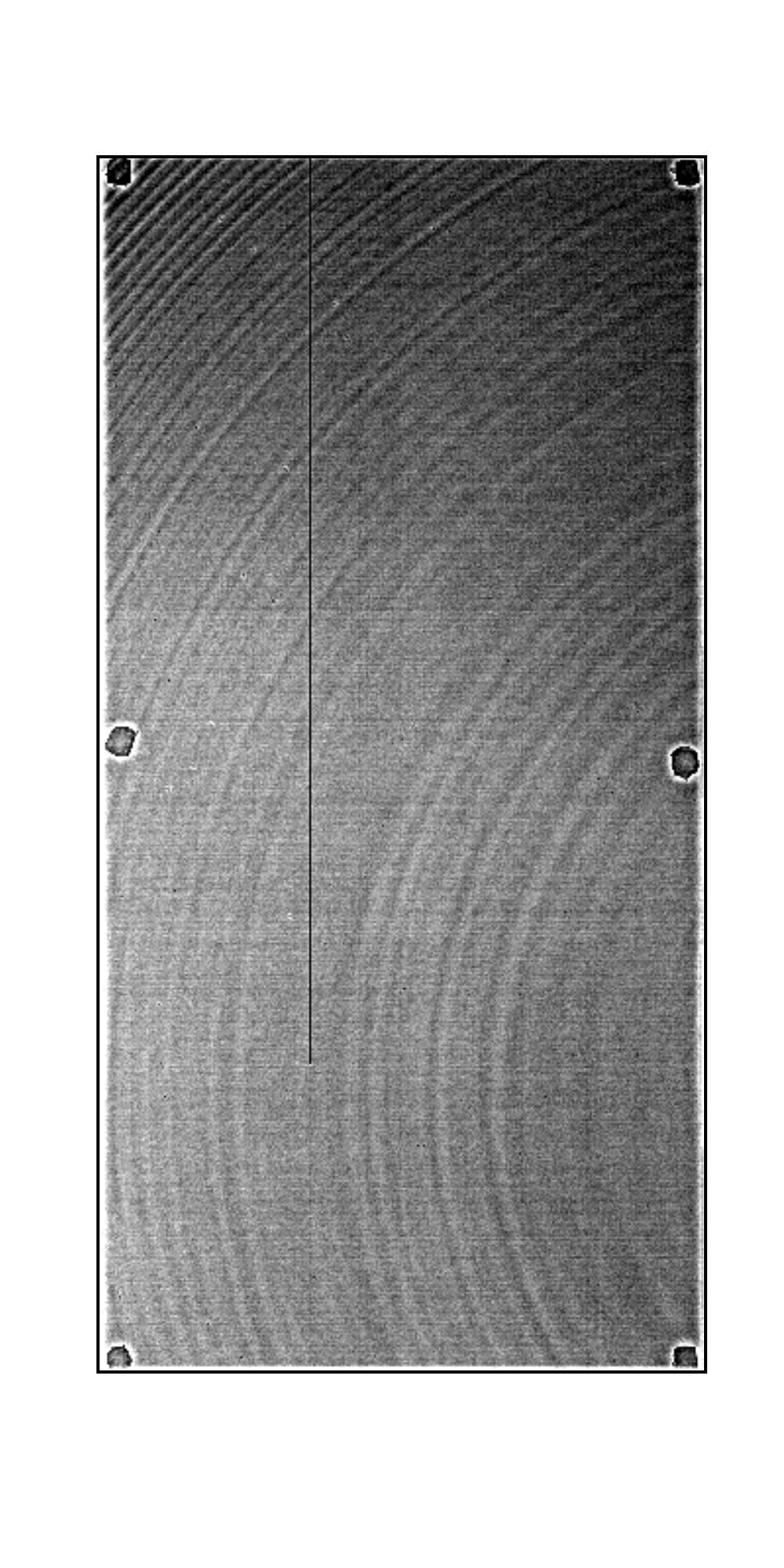}
\hspace{1in}
\includegraphics[height=4in]{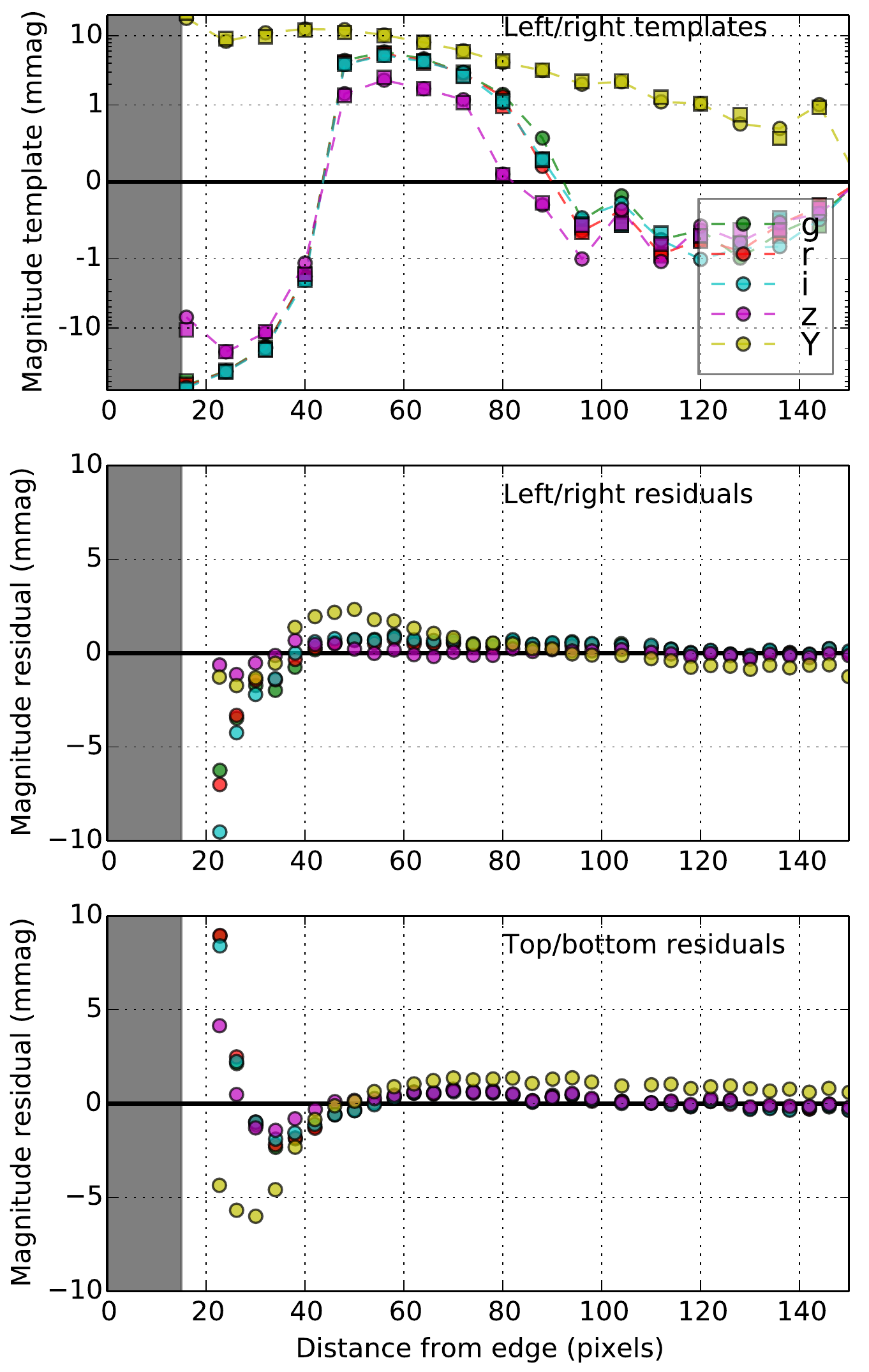}
\end{center}
\caption[]{\small At left is a representative dome flat (CCD N12, $i$-band).
  Visible behavior is dominated by concentric ``tree rings,'' a
  brightening at the edges of the device, and 6 spots where there are
  spacers between the CCD and its carrier.  All of these features are
  primarily pixel-area variations, not QE variations.  We model the
  photometric effect of the edges and rings, and flag the spots as
  regions of less-reliable photometry.  The grayscale spans a contrast
  of 5\%.  At right top are empirical models of the photometric errors induced
  by the edge effect along the long edges, averaged over all CCDs
  (note the log scale).  At middle right, the residual photometric
  error vs distance from the long edge after the empirical model has
  been subtracted from the star flat photometry.  At bottom right is
  the error signal along the short (serial-register) edges, which we
  do not bother to correct.  The 15 pixels nearest each edge are
  masked in processing, and edge effects are reduced to well below
  1~mmag at $\ge40$ pixels from the edge, except in $Y$ band.
}
\label{domeflat}
\end{figure}

Figure~\ref{domeflat} shows a dome flat image of a representative
DECam CCD.  The most prominent features are a series of concentric
rings, and a substantial brightening near the edges of the device.  As
investigated in detail by \citet{andres}, these features are
variations in pixel solid angle \vOmega\ due to stray electric fields
in the detector---they are not variations in the response function
\vrr\ of the sensor material.  As such it becomes the star flat's job to \emph{remove} these
features from dome-flattened images to recover correct aperture
photometry.  We exploit the distinctive geometries of these effects to
produce models for each.  For the tree rings, we average the dome
flats in bins of radius around the apparent center of the rings, and
use this as the lookup function for a radial \texttt{Template}
photomap.  

For the edge functions, we first include a free
\texttt{Piecewise} function of $x$ pixel coordinate, with nodes 
spaced at $\Delta x=8$~pixels for $16\le x \le 144$ and $1905\le
x \le 2033$.  The 15 pixels nearest all device edges are ignored in
DECam processing, as the edge distortions are
slightly dependent on illumination level and too large for reliable 
correction to the desired accuracy.
This means that 6\arcsec\ apertures are incomplete for stars centered
$<26$ pixels from any edge, and therefore the photometry corrections that we
derive from this aperture photometry is decreasingly reliable within
this range.  During standard DES processing, objects intruding into this region are
flagged as having less reliable photometry \citep{detrend,desdm}.
 
We perform an initial round of $\chi^2$ optimization
that leaves all of the nodal values as
free parameters (30 per CCD), fitting simultaneously to
many epochs of star flats to build up stellar samples.
After interpolating over indeterminate nodal values caused by column
defects on some CCDs, these lookup tables are converted to
\texttt{Template} photomaps with fixed shape for further rounds of
fitting.  

See \citet{andres} and \citet{decamast} for deeper discussions of
these device characteristics. Figures~\ref{domeflat} demonstrate that
residual edge photometric errors after our corrections are well below
1~mmag, except within a $\approx20$ pixel strip just inside the masked
region (\ie\ within $\approx35$ pixels of the edge), and slightly larger in $Y$ band.  The only mystery in this
process is that the best-fitting scaling for the tree-ring templates
are $s\approx 0.9$ rather than the expected $s=1$, i.e. the
spurious stellar photometric signature seems slightly smaller than the
fluctuations in the dome flats.

\subsubsection{Optics/CCD polynomials}
The largest difference between the dome flat and the correct response
function \vrr\ is due to contamination of the former by stray
light, which we expect to vary smoothly across the FOV.  We
expect other significant contribution to the star flat correction
$S(\vx)$ from the color difference between the dome-flat lamps and the
reference stellar spectrum, which in turn depends on the spectral response of
each CCD.  The DECam photometric model incorporates an independent
cubic polynomial per CCD to fit these effects.  

\begin{figure}[!t]
\plotone{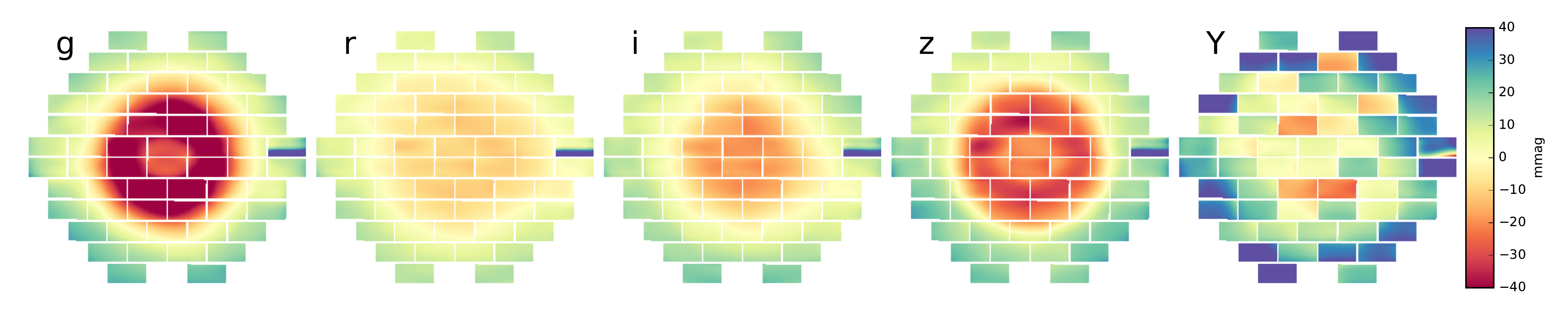}
\plotone{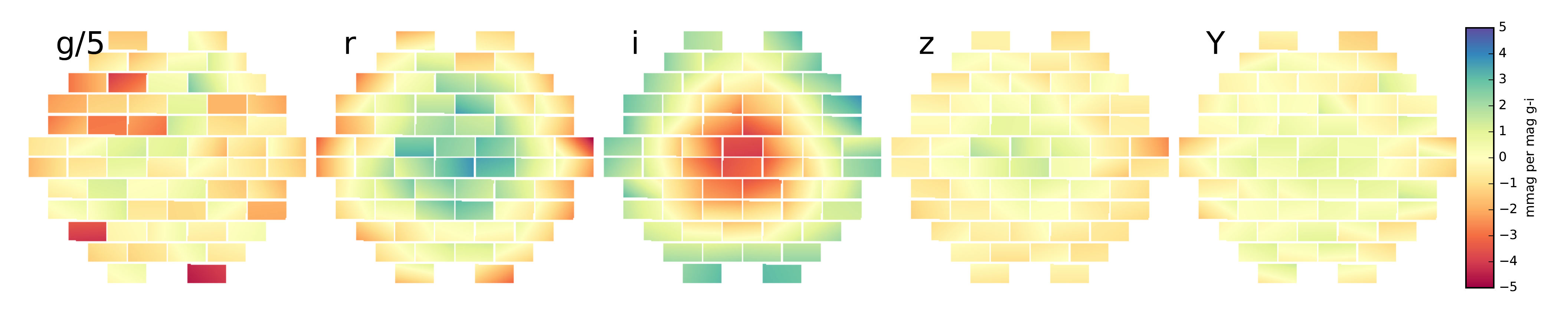}
\plotone{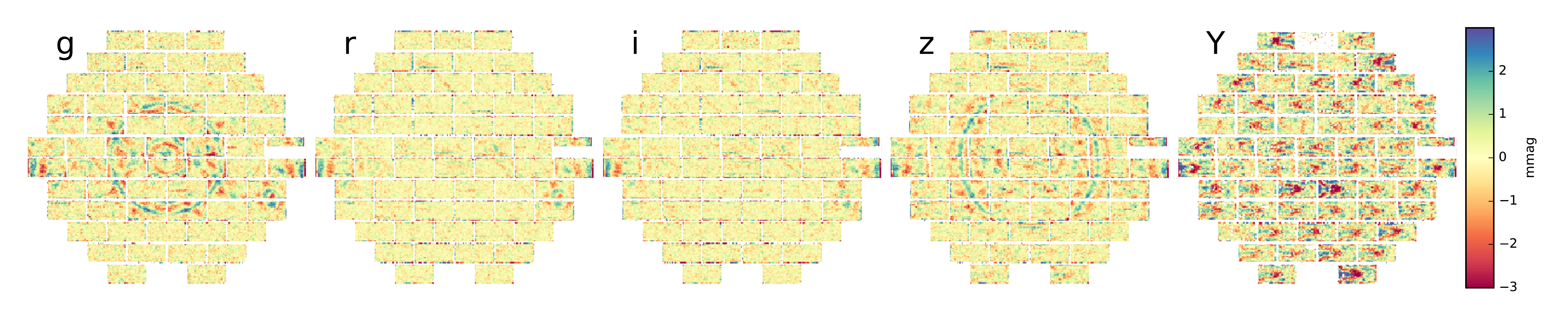}
\caption[]{\small The top and middle rows show the star flat correction
  $S(\vx)$ and color term $c(\vx)$, respectively, for the DECam
  $grizY$ bands, derived from fitting to all star flat epochs
  simultaneously. [Tree ring and edge terms have been suppressed in
  the top row since they are unresolved at the plotted scale.]
  These are the corrections that must be applied to
  stellar photometry of dome-flattened images in order to homogenize
  the photometric response across the array.   The bottom row shows
  the mean photometric residual, binned by array position, for all unclipped
  detections in all of the star flat observing sequences. Only a few
  unmodelled features are visible above 1~mmag level, except in $Y$ band.}
\label{plotflats}
\end{figure}

The top row of Figure~\ref{plotflats} displays the best-fit star flat
corrections $S(\vx)$ as 
a function of position across the array in the $grizY$ filters.  The
dominant feature in the $g$ and $z$ bands is a ``doughnut'' which is
roughly in the position predicted by models of stray light reflections
from the CCD and corrector-lens surfaces.  The star flat correction to
the dome flat is in excess of 50~mmag peak-to-peak, implying that
un-focussed photons
comprise at least $5\%$ of the dome flat flux in places.  No doughnut
is visible in $r, i,$ or $Y$ bands.  In $r$ and $i$, the star flat
correction is $\approx \pm 10$~mmag in a radial gradient that
resembles these filters' color terms.
In $Y$ band we see, in addition, steps between CCDs, 
readily attributed to variation in the red-edge QE of the devices coupled with
a color difference between the dome lamps and the stellar reference
spectrum. 

\subsubsection{Color terms}
The color term $c(\vx)$ is also taken to be an independent
polynomial per CCD per filter.  Linear order suffices to
remove any detectable color response patterns.  The middle row of
Figure~\ref{plotflats} shows the best-fit solutions
derived by \photofit.  
In $g$ band (and to a lesser extent
$Y$ band) we can see the consequences of device-to-device variations
in the blue (red) end of the CCD QE spectrum, leading to color terms
as large as $\approx 25$~mmag per mag change in $g-i$.
Variation from the natural
passband is smaller in the other bands, at $\lesssim5$~mmag/mag.
The
mild radial gradient in $i$ band is shown by \citet{ting} to be
consistent with the known radial gradient in the blue-side cutoff
wavelength of the filter.    

\subsubsection{Extinction gradient}
Atmospheric extinction will vary across the 1\arcdeg\ radius of the DECam FOV.
If the zenith angle is $z$, the airmass at the
optic axis zenith angle $z_0$ is $X=\sec
z_0$, and the extinction is $k X$ for some constant $k$, then the
first-order correction for extinction away from the telescope axis is
\begin{equation}
\Delta m = X\sqrt{X^2-1} \, k\,\Delta z.
\label{extinction}
\end{equation}
We take a nominal extinction constant of $k=(0.2,0.1,0.07,0.08,0.07)$
mag per airmass in $(g,r,i,z,Y)$.  At the maximum field radius $\Delta
z=\pm1\arcdeg,$ \eqq{extinction} grows to a $\pm21$~mmag deviation for
$X=2$ in $g$ band. The extinction gradient must be corrected, even in the
redder bands at more modest airmass, to attain mmag homogeneity.  We do so by precomputing
a \texttt{Polynomial} photomap for each exposure for the
\texttt{gradient} term, which is held constant during
fitting.  In future DES photometric solutions we could account for
temporal variation in the atmospheric extinction constant $k$.

The second derivative of extinction contributes $\Delta m<1$~mmag
in peak-to-peak amplitude even at $X=2$ in $g$ band, so we can
approximate extinction variation with just the linear term.

\subsection{Model residual patterns}
The bottom row of Figure~\ref{plotflats} shows the
residual of the stellar photometry to the \photofit\ solution in each filter, averaged over all
star-flat-sequence detections and binned by position in the DECam
focal plane.  The performance of the model is excellent in $griz$ bands:
the RMS of the means in the 34\arcsec\ square bins is
$<1$~mmag and visually consistent with noise, with the
exception of a few coherent features at 1--2~mmag level.
Residuals are visible at the edges of
$g$ and $z$ band doughnuts, indicating sharper edges
than our polynomials can capture.  There also appear to be features in
the east- and west-most CCDs' response that are higher-order than our
cubic.  

\begin{figure}[tb]
\begin{center}
\includegraphics[width=\textwidth]{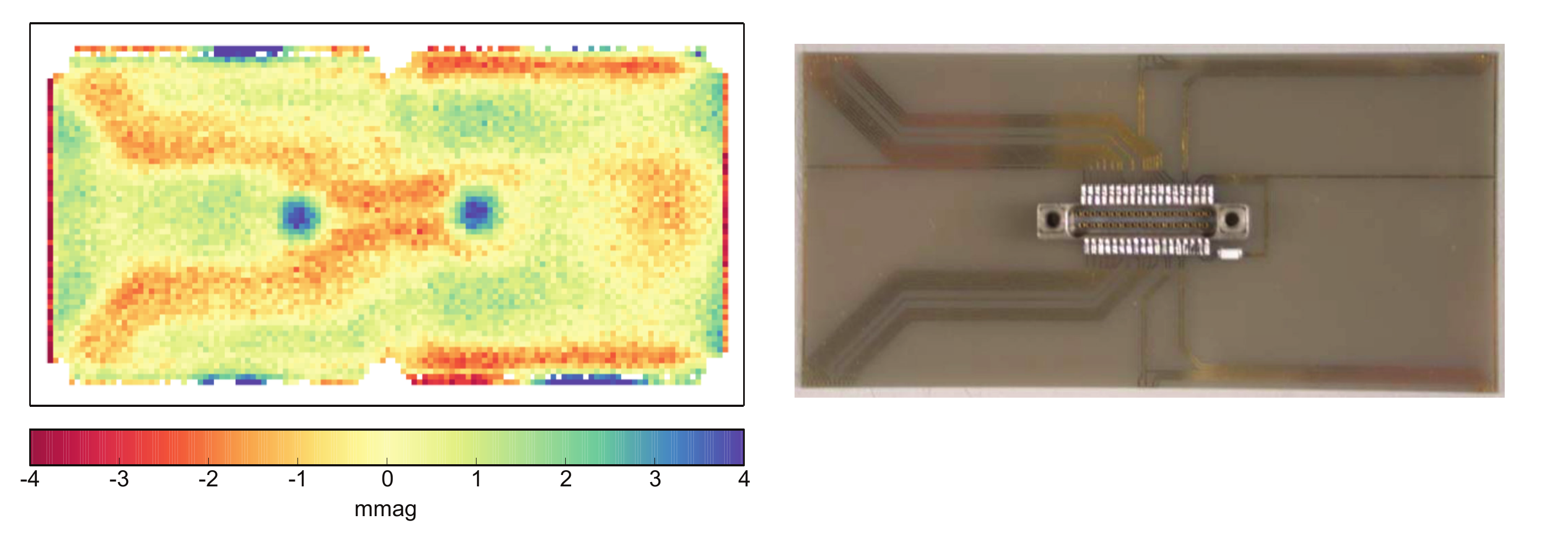}
\end{center}
\caption[]{\small At left: the residual stellar magnitude errors in $Y$ band,
  after star flat correction, binned by position on the CCD.  All CCDs
and all star flat epochs are stacked.  The $\Omega$ shape follows the
traces on the aluminum nitride board to which the devices are mounted, shown at
right (courtesy J. Estrada and T. Diehl).}
\label{yresids}
\end{figure}

The $Y$ band shows significantly larger residual variation
($\approx1$~mmag RMS), which close examination reveals is primarily a
repeating pattern on all devices.  Figure~\ref{yresids} shows the mean
residuals binned by position on the CCD, stacking all devices for all
star flat exposures in $Y$ band. Remarkably we see that the star flat
detects the presence of metallic structures on the board to which the
CCD is mounted.  The red edge of the $Y$ band is defined by silicon
bandgap energy, and in $Y$ band a significant fraction
of stellar photons make it through the CCD to the mounting board, where the
reflectivity of the 
mount will affect QE.  A difference between the dome and 
reference spectra leads to this pattern.  It is present at $\ll1$~mmag
in $z$ band and not detected in $g, r,$ or $i.$  We leave this pattern
uncorrected since precision $Y$-band photometry is not central to DES
science.  

The device-stacked residuals in $griz$ bands reveal weak features along the long edges
which indicate a break in our assumption that the ``glowing edge'' is
constant for all rows. We leave these patterns uncorrected as well, since
they are weak and affect very little of the focal plane.

\subsection{Error statistics}
\label{errorstats}
Our goal is to eliminate any variance in magnitude estimates of stars
in excess of that expected from photon shot noise and detector read
noise.  We define the excess RMS noise $\sigma_{\rm sys}$ as
\begin{align}
\label{excessnoise}
\sigma^2_{\rm sys} & \equiv \left\langle \Delta m_i^2 - \sigma^2_{{\rm
                     stat},i} \right\rangle, \\
\Delta m_i & \equiv \frac{m_i - \bar m}{\sqrt{1-w_i/\sum w_j }}, \\
w_i & \equiv \sigma_{{\rm stat},i}^{-2}, \\
\bar m & \equiv \frac{\sum w_j m_j}{\sum w_j}.
\end{align}
This $\sigma_{\rm sys}$ is distinct from the value used to set an
error floor in \eqq{chistar}, though the value measured here is a good
choice for setting the error floor.  The $\sum w_j$ terms are over all
measures of the same star.
Note that $\Delta m_i$ is normalized such that the expectation of
$\sigma^2_{\rm sys}$ is zero if each measurement $m_i$ has
variance equal to its expected statistical noise
$\sigma_{{\rm stat},i}$ and all measurements have independent errors.
\textsc{SExtractor} provides an estimate of $\sigma_{{\rm stat},i}$
expected from the photon shot noise and read noise within the stellar
aperture. 

\begin{figure}[tb]
\plottwo{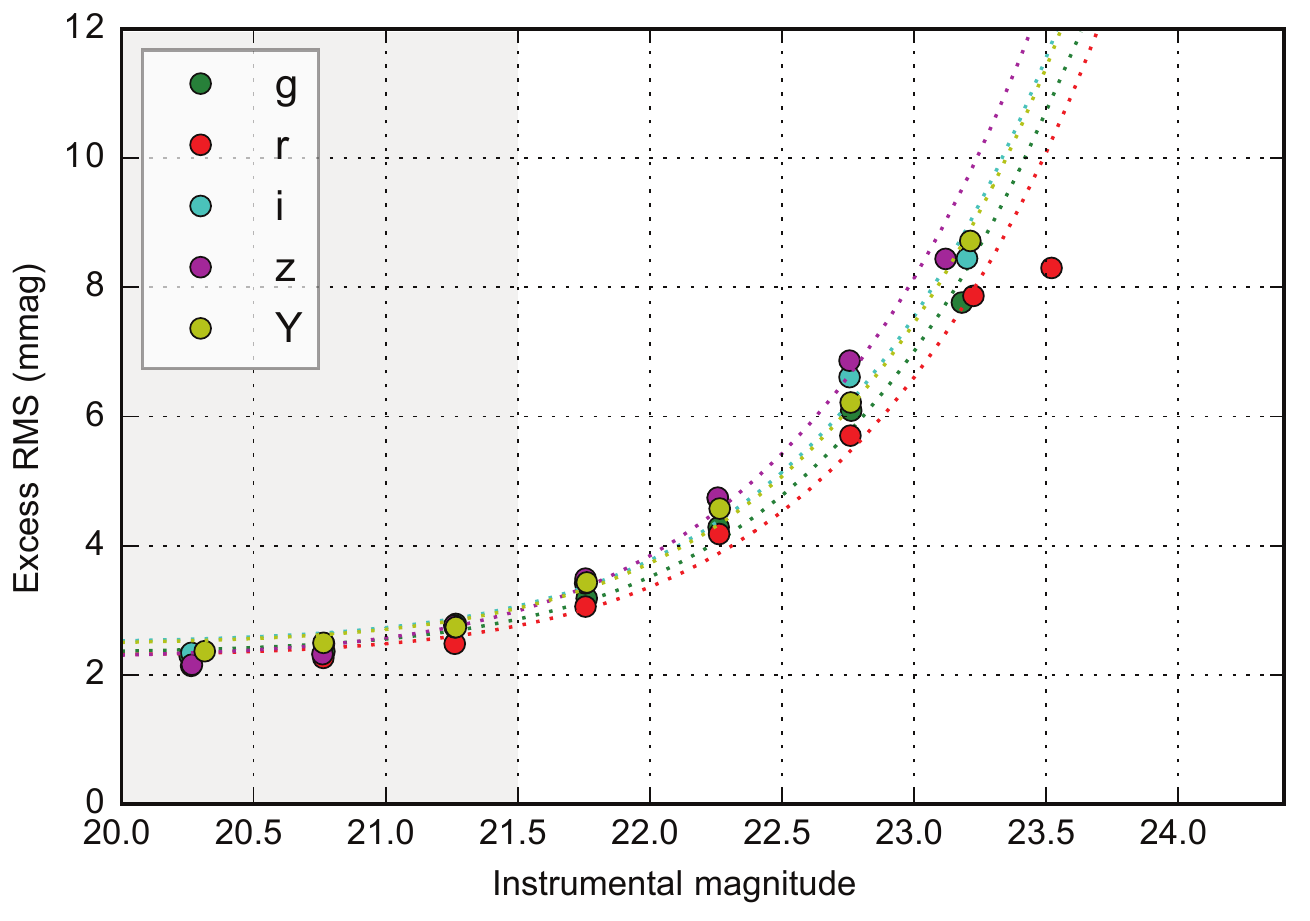}{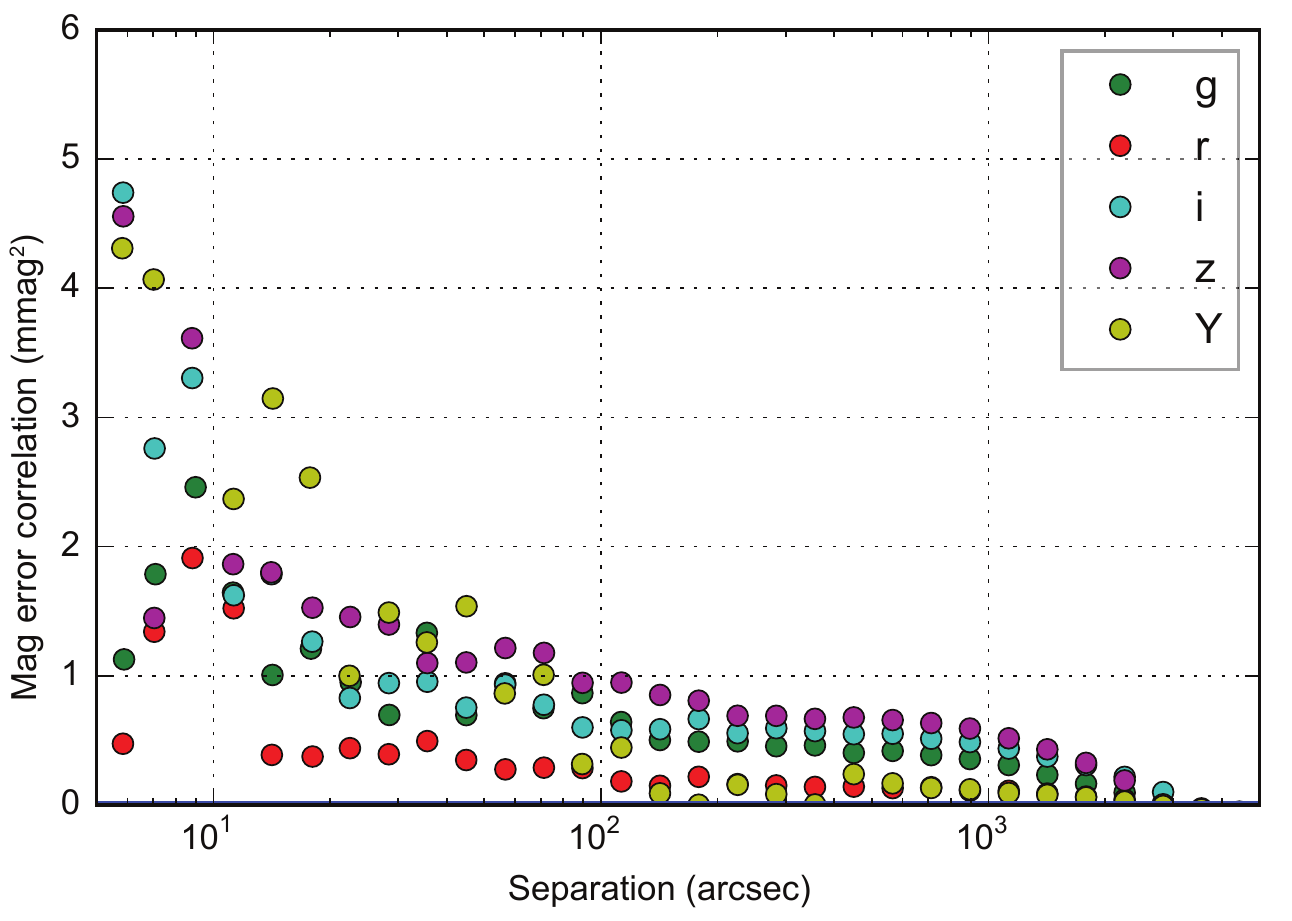}
\caption[]{\small Left: the RMS excess magnitude noise, defined by
  \eqq{excessnoise}, is plotted against instrumental magnitude for
  exposures from the 20131115 star flat sequence for each filter.  
  The dotted lines show fits to the model of
  \eqq{excessmodel}.  There is a 2--3~mmag floor on the magnitude errors
  above statistical noise. Further statistical analyses make use of
  stars in the shade range of magnitudes.  At right is the 2-point correlation
  function of magnitude residuals for the same data.  In
  $griz$ bands, any photometric errors correlated at separations
  $\ge20\arcsec$ are $<1$~mmag.}
\label{magrms}
\end{figure}

The left panel of Figure~\ref{magrms} plots the excess RMS noise vs
stellar instrumental magnitude for the 20131115 star flat data.  The
data are consistent with the dotted curves for the model
\begin{equation}
\label{excessmodel}
\sigma^2_{\rm sys} = \sigma^2_m + \left(\frac{2.5}{\log 10}\right)^2
\left(\frac{\sigma_f}{f}\right)^2.
\end{equation}
The free parameters of this model are: $\sigma_m$, an error that is
fixed in magnitudes and hence represents a multiplicative error in
flux as would be expected from flat-fielding imperfections; and
$\sigma_f$, an error that is fixed in flux units.

We believe that the non-zero fixed-flux error $\sigma_f,$ is
attributable to noise in the estimation of the sky level due to shot
noise in the annulus used for local sky determination.  \sextractor\
does not include such noise in its error estimation, and we find that
propagating the expected sky-annulus mean error into flux measures in
our large (6\arcsec) apertures yields roughly the right level of $\sigma_f$.
Reducing the size of the photometric aperture from 6\arcsec\ to
4\arcsec\ diameter reduces $\sigma_f$ by very close to the ratio
$(4/6)^2$ of aperture area, supporting this hypothesis.  
Henceforth we will concentrate analysis of the residual errors on
stars with instrumental magnitude $20<m_i<21.5$, corresponding to
$>10^4\,e$/s flux, where the effects of $\sigma_f$ are subdominant to
the calibration errors $\sigma_m$ that are our primary interest. 

The derived values of $\sigma_m$ are in the range 2--3~mmag for most
of the star-flat exposures.  \emph{The star flat solutions for a given night
  homogenize the array calibration to 2--3~mmag RMS.}  The level of
excess variance does change over time---some nights or exposures are
worse, as we will investigate below.

More indicative of photometric calibration errors is
the 2-point angular
correlation function (2PCF) $\xi_m(\theta)$ of the magnitude residuals $\Delta m_i$.
We consider only measurement pairs within the same exposure.  Shot
noise within the stellar aperture will not contribute to the 2PCF, nor
will stellar variability, and
shot noise within the sky annuli will not contribute beyond the
$\approx20\arcsec$ diameter of the \textsc{SExtractor} local sky
annuli.  As seen
the right panel of Figure~\ref{magrms}, the 2PCF amplitude at
$\theta>20\arcsec$ separation \emph{is typically below 1~mmag$^2$ in $griz$
  bands.} The slightly higher level in $Y$ band is likely due to
the uncorrected static pattern detected in Figure~\ref{yresids}.
The level of correlated noise at $\theta>20\arcsec$ does exceed 1~mmag in
some exposures or consistently on some nights, which will be
characterized in Section~\ref{atmosphere}.

We conclude that most of the 2--3~mmag
excess calibration noise arises from effects with short coherence
length.  Stellar variability would produce such a signature, but each
filter's star flats span $<30$~minutes time so this is likely
unimportant for this analysis.
Aside
from the tree rings and glowing edges, our star flat models do not
attempt to correct small-scale error in the dome flats.  A close look
at the dome flats reveals pixel-to-pixel variations that repeat from
band to band.  These are probably pixel-area variations from
imperfections in lithography of the CCD gate structure.  If we ascribe
\emph{all} the small-scale variation in the dome flats to pixel-area
rather than QE changes, we estimate that in seeing of 1\arcsec\ FWHM,
these flat-field errors induce $\sigma_m\approx 1.2$~mmag. In worse
seeing, $\sigma_m$ decreases as the signal spreads over more pixels
and averages down the gate errors.  These single-pixel-scale errors
are essentially impossible to characterize with on-sky measurements.
Future experiments, such as LSST,
would benefit from laboratory characterization of lithographic errors
if it is desired to reduce excess RMS to $<1$~mmag.
We note, though, that for DES measurements such as galaxy clustering, 
bias is induced only by calibration errors that correlate between
targets, and we have demonstrated that these can be reduced to
$\approx 1$~mmag for time scales $<1$~hour.

\section{Short-term stability}
\label{atmosphere}
For each star flat epoch, we have fit a fixed solution to all
22 exposures in a given filter.  For the analyses in this section, the
only additional degree of freedom in an individual exposure's solution
is an overall \texttt{Constant} photomap to allow for zeropoint
variation.  Recall that we also have a term per exposure for
extinction gradient, but this is held fixed to an \textit{a priori} model.
We now look for patterns in the deviations of individual exposures'
photometry from the night's solution, \ie\ is there detectable
variation in photometric response pattern within the $\approx25$~min
it takes to expose a given filter?

\begin{figure}[!t]
\plotone{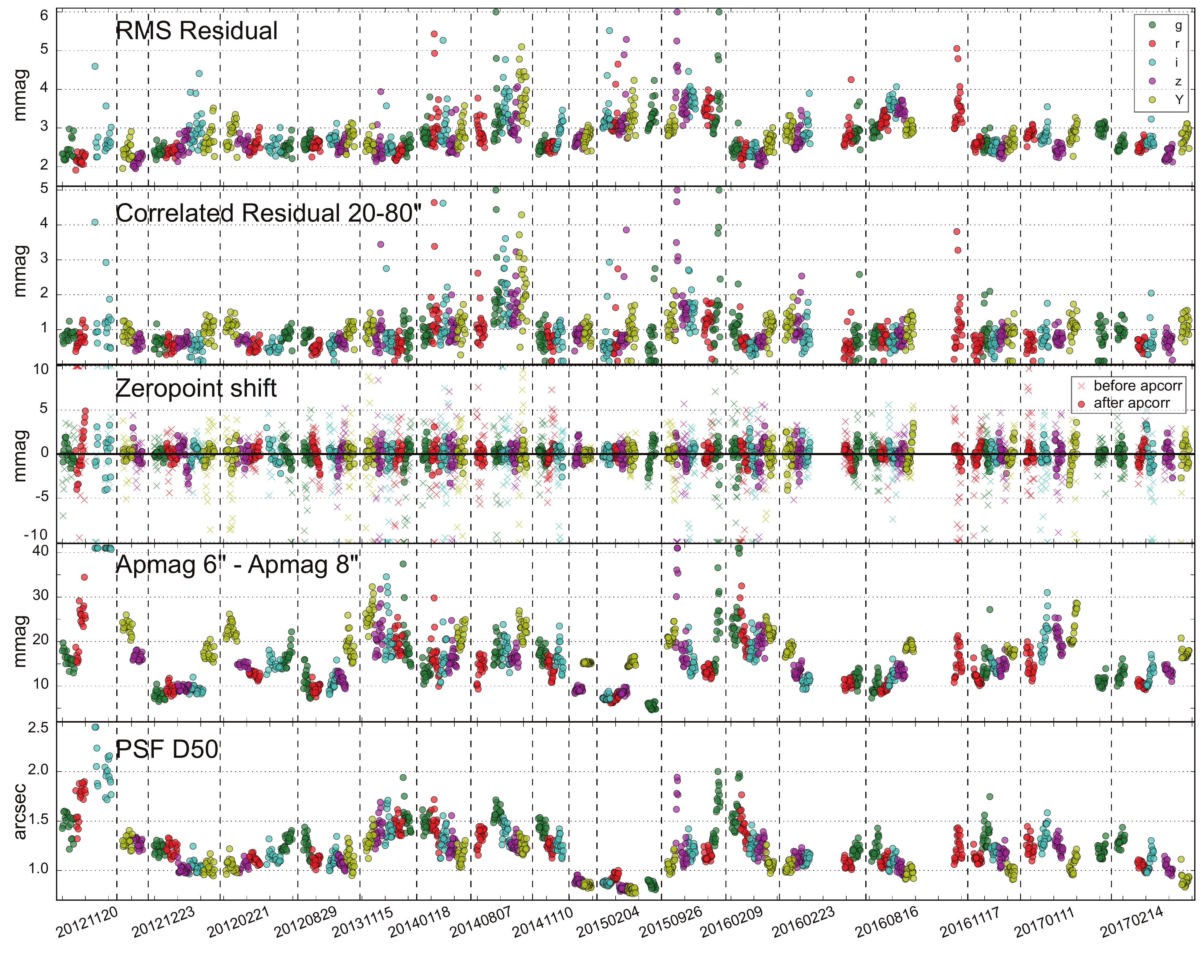}
\caption[]{\small Each panel shows a photometry-related variable for each of
  the $\sim1800$ exposures taken during the 16 cloudless star-flat
  observing epochs.  The tick marks on the horizontal axis mark 1-hour
  intervals, but the vertical dashed lines represent breaks of days to
  months.  For this plot, each star flat epoch has its own photometric solution.
The top row plots the RMS photometric noise above the expected shot
noise for stars with instrumental magnitude $20<m_{\rm inst}<21.5,$
for which the errors from sky-level determination are usually negligible.
Most nights this is 2--3~mmag, but some nights and individual
exposures are worse.  The second row plots the RMS amplitude of
correlated magnitude errors,
$\sqrt{\xi_m(20\arcsec<\theta<80\arcsec)},$ which is seen to be
  $\approx 1$~mmag in $griz$, again with a few nights 
  and a few stray exposures of higher level.
The third row plots the deviation of the exposure zeropoint shift
from the atmospheric secant law for that night.  The crosses
give the initial values, which are 2--5~mmag RMS, but when the
aperture correction of Section~\ref{apcorr} is applied, the RMS
zeropoint variation is reduced to $\approx1$~mmag.  The fourth row
shows the aperture correction proxy $A_t$, the difference between
8\arcsec\ and 6\arcsec\ aperture magnitudes.  The fifth row plots the
half-light diameter $D_{50}$ of the PSF, which is seen to be less than
fully predictive of $A_t$ or the zeropoint shifts.}
\label{timeplots}
\end{figure}

Figures~\ref{timeplots} plot the primary diagnostics for unmodelled
time variation in photometric response.  Of most interest is the
second row: we plot the amplitude of coherent magnitude residuals 
$\sqrt{\xi_m(\theta)}$ using a broad bin $20\arcsec<\theta<80\arcsec$
(see Figure~\ref{magrms}).  Usually $\sqrt{\xi_m(\theta)}\le 1$~mmag
(except in $Y$ band)---excursions above 2~mmag indicate the
presence of changes in the response pattern.
Additional evidence of varying instrument response is
given by elevated RMS residual of bright stars, plotted in the top row,
although keep in mind that the RMS can also be inflated by errors in
sky background determination, which are not of interest in this paper.

\subsection{Freaks}
\begin{figure}
\plotone{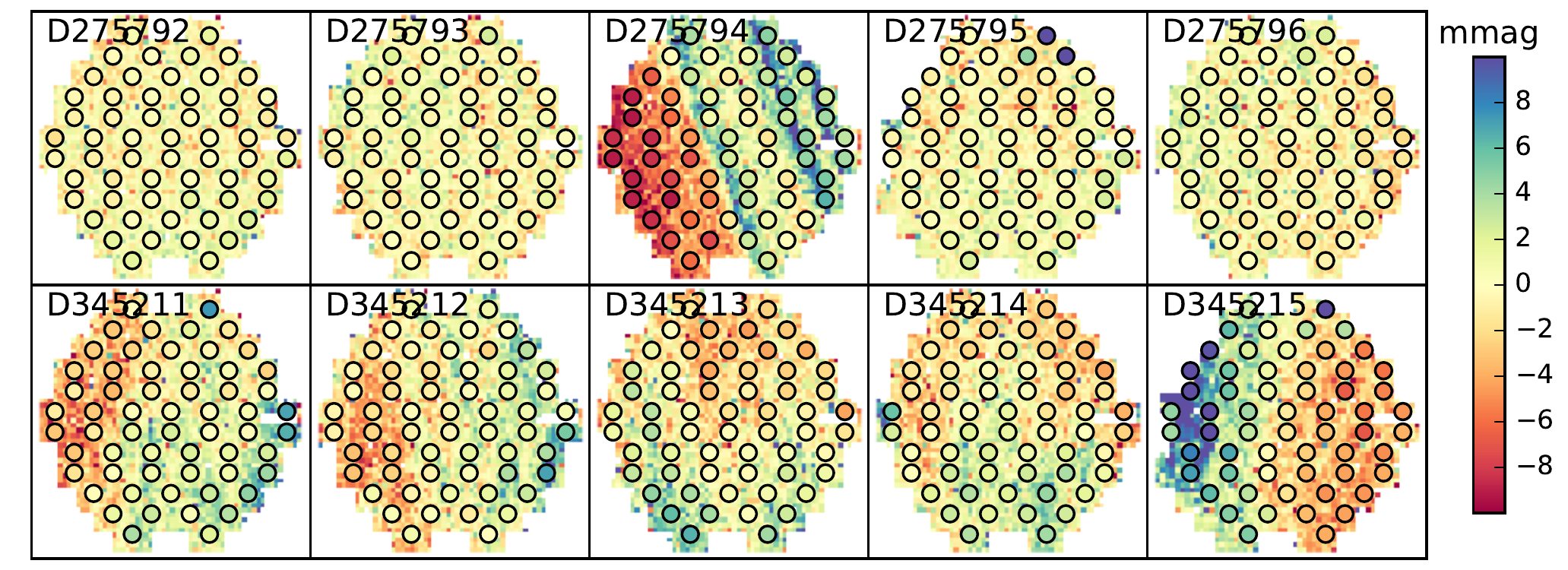}
\caption[]{\small The background of each panel plots the spatially
  binned residual photometric errors of bright stars in a single
  exposure, after application of the static starflat and color
  corrections $S(\vx)$ and $c(\vx)$.  The colors of the overplotted
  circles encode $2\times$ the variation of the aperture correction proxy
  $A_{ti} $ (equation~\ref{apcorrdef}) across the FOV.  It might be
  difficult to notice that the colors inside the circles are different
  from the color just outside the circles, which tells us that
  the variation in $A_{ti}$ is a very accurate
  predictor of the spatial pattern of photometric errors.  The top row
  shows a series of star flat exposures at 1 minute interval
  containing one freak excursion.  The lower row shows a sequence of
  exposures during a period of very unstable seeing.  In all cases the
  photometric inhomogeneity is clearly attributable to variations in
  the fraction of flux falling outside the nominal 6\arcsec\
  photometric aperture.
}
\label{freaks}
\end{figure}

There are occasional isolated instances of higher photometric
residuals in a single exposure, for example \#275794 taken 
during the 20131115 $i$-band star flat sequence.
The top row of Figure~\ref{freaks} plots the photometric residuals to
the static model observed on a series of 5 consecutive exposures.
Exposure 275794 exhibits large coherent residuals of $\pm10$~mmag,
while exposures taken just 1 minute earlier or later have the usual
$\approx1$~mmag RMS residual.  

We find strong evidence that these
residuals are due to spatial fluctuations in the \emph{aperture
  correction} rather than in the \emph{transparency} of the
atmosphere, as follows.  Define $A_{ti}$ to be the median differences
between 8\arcsec\ and 6\arcsec\ aperture magnitudes of bright stars on
CCD $i$ in exposure $t$, as per \eqq{apcorrdef}.  The color of 
each circle overplotted in Figure~\ref{freaks} indicates $2(A_{ti} - \langle A_{ti}\rangle)$
for each of the 60 CCDs in use for these exposures.
This scaling of the aperture-correction proxy $A_{ti}$ is seen to be a
near-perfect match 
to the pattern of photometric residuals across the FOV for this
exposure.  From this we conclude that (a) the excess correlated
photometric errors in this exposure are caused by variations in
aperture correction across the FOV, and (b) the statistic $A_{ti}$ is a
good tracer of such variations.

What caused this freak exposure on an otherwise well-behaved,
cloudless night?  The streaky appearance of the residuals is likely
the result of some disturbance being blown across the field of view
during the exposure.  We do not have any well-supported theories for
what kind of disturbance would cause a localized increase in
the PSF wing strength.  The number of outlier points in the second row of
Figure~\ref{timeplots} gives some indication that $O(1\%)$ of
exposures have freak disturbances, in addition to there being some
nights with consistently high correlated residuals.  One possibility,
as yet unverified, is that these are associated with elevated levels
of scattering or turbulence in airplane contrails.

\subsection{Noisy nights}
On some nights $\xi_m$ is consistently or frequently well above the
typical 1~mmag$^2$ 
level, \ie\ on the nights 20140807 and 20150926.  The lower panel of
Figure~\ref{freaks} shows the magnitude residuals for a series of 5
consecutive $z$-band exposures on the latter night, during which the
seeing half-light diameter underwent a rapid excursion from  1\farcs3
to 2\farcs0 and back (see bottom row of Figure~\ref{timeplots}).  The
overplotted circles in Figure~\ref{freaks} once again indicates that
the photometric errors are closely tracked by variations in the
aperture correction proxy $A_t$ across the focal plane.

Indeed it is found that nearly every star-flat exposure $t$ with high photometric
residual $\sqrt{\xi_m}$ is accompanied by higher-than average
dispersion of the $A_{ti}$ across the focal plane.

\subsection{Zeropoint stability}
\label{apcorr}
A basic question of critical importance to ground-based photometric
survey calibration is:  just how stable is the atmospheric
transmission on a given night, apart from expected scaling with $\sec
z$? We answer this question by calculating the deviation between each
exposure's zeropoint and the best-fitting secant law, as in the
numerator of \eqq{atmprior}.  We adopt $\sigma_n=0.5$~mmag in our
atmospheric prior to give the $\chi^2$ minimization strong incentive
to reduce the zeropoint residual to mmag level.

\begin{deluxetable}{lccccc}
\tablewidth{0pt}
\tablecolumns{6}
\tablecaption{Zeropoint variation per filter}
\tablehead{
\colhead{Band} &
\colhead{$g$} &
\colhead{$r$} &
\colhead{$i$} &
\colhead{$z$} &
\colhead{$Y$}  
}
\startdata
RMS variation before aperture correction (mmag) & 2.6 & 3.1 & 4.5 &
1.9 & 2.9 \\
RMS variation after aperture correction (mmag) & 0.9 & 0.7 & 0.8 & 1.1
& 1.0 \\
Nominal apcorr coefficient $k_{n3}$ & 1.6 & 1.9 & 1.9 & 1.9 & 2.2 \\
\enddata
\label{zptable}
\end{deluxetable}

The crosses in the third row of Figure~\ref{timeplots} plots these
residuals for each star flat exposure before we introduce any aperture
corrections, \ie\ we set the aperture correction coefficient $k_{n3}=0$ in
\eqq{atmo1}.  The RMS values in each filter, shown
in the first row of Table~\ref{zptable}, are 2--5~mmag.  There is
substantial variation in RMS from night to night, and deviations
$>10$~mmag in individual exposures are common.

The temporal zeropoint jitter is highly
correlated with estimators of the fraction of light falling
outside the photometric aperture.  The fourth and fifth rows of
Figure~\ref{timeplots} show two variables we might expect to correlate
with the aperture correction, namely the half-light diameter of the
PSF (bottom row) and the $A_t$ defined in \eqq{apcorrdef} as the
fraction of extra light found in extending the aperture from 6\arcsec\
to 8\arcsec.  Both quantities are the median of all bright stars
in the field.
The latter variable is found to correlate much better
with the zeropoint jitter.  $A_t$ does not measure the aperture
correction to infinity, but it is sensible to think that the aperture
correction to 8\arcsec\ would correlate with the correction to
infinity, so we introduce the term $k_{n3}\times A_t$ into the
zeropoint model as per \eqq{atmprior} essentially as an adjustment to
each exposure's zeropoint.  Leaving $k_{n3}$ as a free
parameter on each night reduces the zeropoint jitter to $\le 1$~mmag
RMS ($z$ band is slightly higher, perhaps due to some variability in
water vapor absorption).  

We find that setting $k_{n3}$ to a nominal value in each filter
yields zeropoint jitter indistinguishable from having a free value for
each night.  The nominal $k_{n3}$ values and the resultant RMS values
per filter are in Table~\ref{zptable}.   The circles in the 3rd row of
Figure~\ref{timeplots} show the truly impressive stability in
atmospheric transmission obtained after making the zeropoint
correction.

In summary, we find that \emph{all} of the deviations above $\approx
1$~mmag RMS from a static
response function plus secant airmass law on short timescales are
plausibly attributable to spatial/temporal variations in aperture
corrections.  The $A_t$ statistic measured from bright stars is an
accurate predictor of these aperture corrections, so on a typical
half-hour stretch of clear-sky observations we can homogenize
the exposure zeropoints to $\approx 1$~mmag, and if we have sufficient
stellar data in an exposure to map out variation of $A_t$ across the
FOV, we could reduce any intra-exposure inhomogeneity to similar
level.

\section{Long-term stability}
\label{stability}
To investigate the stability of the reference response $\vrr(t)$ over
months to years, we fit the entire multi-year ensemble of star flat
data to a common instrument model and examine the mean residuals to
this model in each epoch (recall they have all be flattened
with a common dome flat too).  We observe changes of several mmag
between star flat epochs, smoothly varying across the focal plane, and
very similar in all filters.  This ``gray'' term might arise from
accumulation of dust or contaminants on one of the lenses (or the
detectors).  

The $Y$ band exhibits additional shifts in the response of
specific CCDs.  After cycling the camera to room
temperature, the strength of some thermal contacts change slightly and
the equilibrium temperatures of the devices change slightly.  It is
thought that these temperature shifts will then change the silicon
optical depth near the band gap and slightly alter the $Y$-band
response.  

Whenever we jointly fit data from multiple
  observing nights, we allow a constant \texttt{Color} photomap for
  each exposure, as it is expected that variations in atmospheric
  constituents will change the color response of the system by up to
  several mmag \citep{ting,fgcm}.

\subsection{Test of gray drift model}
To test the hypothesis that long-term changes in instrument response
are due to low-order gray absorption plus CCD-specific changes in
$Y$-band response, we derive a new global response
model as follows:
\begin{enumerate}
\item Run \photofit\ jointly on the $i$ and $z$-band exposures from
  all star flat epochs,
  adding an additional 4th-order \texttt{Polynomial} photomap to the
  model.  Each epoch (except one reference epoch) is allowed an
  independent gray  \texttt{drift} term that is common to both
  filters.  
\item Run \photofit\ on the $g$ filter's data, augmenting the model
with a \texttt{drift} term for each epoch whose parameters are held
fixed to the values derived in step (1).  Repeat for $r,$ $i,$ and $z$
bands.  
\item Run \photofit\ on the $Y$ data, including the fixed
  \texttt{drift} terms plus an additional free \texttt{Constant} for each
  CCD (except one) in each epoch, to track the temperature changes.
\end{enumerate}

\begin{figure}[!p]
\center
\includegraphics[width=7.0in,angle=-90,origin=c]{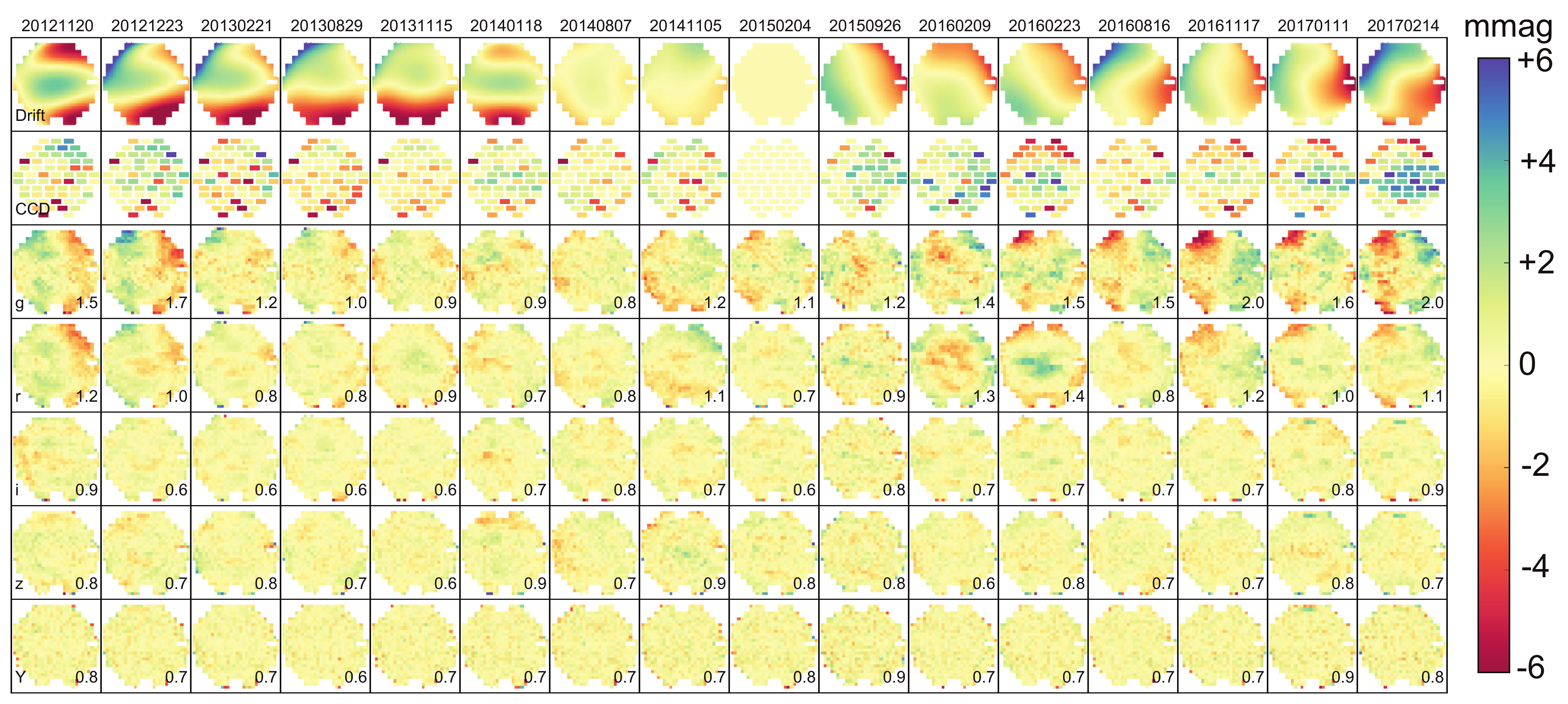}
\caption[]{\small The top row shows the 4th-order polynomial ``drift''
  for each
  epoch that best brings its $iz$ photometric response into agreement
  with epoch 20150204.  The second row shows the shifts of CCD
  response that are further applied to homogenize the $Y$-band
  photometry.  Further rows show the mean photometric error in
  1024-pixel bins at each epoch and filter, to a model including the 
  gray polynomial drift terms (and CCD shifts, for $Y$ band).  The
  number in lower-right of each panel is the RMS of the binned
  residuals (in mmag). A common polynomial in all bands captures the
  epoch-to-epoch changes in response to $<1$~mmag RMS in the $iz$
  bands, but an unmodelled bluer component is apparent in the
  $g$-band residuals with a weaker $r$-band signature. } 
\label{drifts}
\end{figure}

Figure~\ref{drifts} plots the results of including the drift terms in
a fit to more than 4 years' star flat data.  The top row shows the
derived gray polynomial terms, which exhibit amplitude up to
$\pm6$~mmag and RMS up to 2.5~mmag.  The second row plots the $Y$-band
CCD shift terms per epoch, which are as large $\pm 12$~mmag for the
worst devices in the worst epochs.

The remaining rows plot the mean photometric residual to the gray
drift model
across the focal plane for each filter and each epoch.  The RMS
residual signal is below 1~mmag RMS (some of which is measurement
noise and stellar variability) at all times in $izY$ bands.  The gray
model is seen to be not quite sufficient: there appears to be an
additional blue drift component, present in $g$ band with a fainter version
in $r$ band.  This pushes the RMS residual to the gray model up to
2~mmag in the worst $g$-band epochs.  We note that even if this
$g$-band deviation were not corrected, the 2~mmag RMS photometric
variation would be well below DES requirements and far better than any
previous survey's photometric calibration accuracy.

\subsection{Time history of drifts}
The star flat sequences are taken too infrequently to resolve the
time scale for instrumental response changes.  DES observes its
supernova fields roughly once per week during the observing
seasons.
The SN exposures are taken with minimal dithering, so that the same stars
are on a given CCD in every exposure.  This makes the data useless for
determining the spatial structure of the response function $\vrr(t)$,
but valuable for examination of its temporal structure at finer
resolution.  We will use the $z$-band data from field SN-C3, for which
an observing sequence comprises $11\times330$~s exposures.
Roughly 2000 stars are available at $S/N\gtrsim30$, are not saturated, and
have well-determined $g-i$ colors falling within the calibratable range.
Through October 2016, there are 72 nights of SN-C3 $z$-band observations for which there is no
evidence of clouds and the mean seeing half-light diameter is
$<1\farcs6.$  

Images for these exposures were processed using a fixed dome flat for
all 4 years' data, and the \photofit\ assumed a fixed instrumental
response model, with parameters fixed to those determined from the
star flat observations.  Each SN-C3 exposure is given a free zeropoint
and color term, and an extinction gradient correction determined
\textit{a priori.}  The atmospheric prior for each night has fixed
values $k_{n1}=0.08$ for the airmass term and $k_{n3}=1.5$ for the
aperture correction coefficient.  After the fit we calculate the
residual deviation of the stellar instrumental magnitudes from the
\photofit\ model.
Our analysis tracks
variation of the camera response using the median
residual deviations $\Delta m_{ni}$ of the stellar detections
in CCD $i$ on night $n.$

\begin{figure}[!t]
\plotone{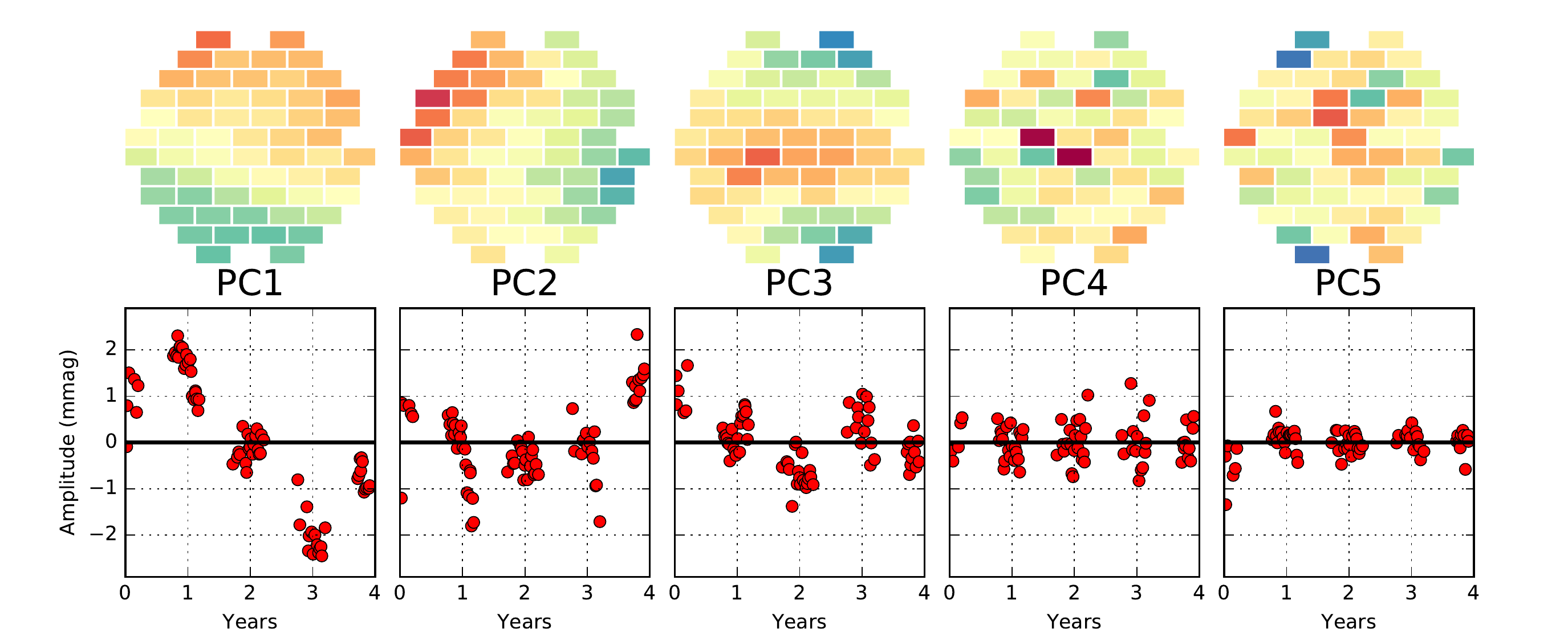}
\caption[]{\small First 5 principal components of variation in the
  response of DECam CCDs to the stars in the SNC3 field.  The top row shows
  the pattern of principal components---the first three are smooth
  functions of array coordinates.  The bottom row plots the
  coefficient of each PC for each night of observations; the PCs are
  normalized such that these coefficients are in units of the RMS
  fluctuation across the focal plane contributed by the PC.  The first
  three are seen to vary smoothly with time over the first 4 years of
  the survey.  The fourth PC does not (but does correlate with
  seeing).  The fourth and higher PCs contribute less than 1~mmag RMS
  photometric error.}
\label{snc3}
\end{figure}

Changes in the photometric residual pattern are observed to be smooth
in both time and in position on the array.
To quantify this, we identify the principal components (PCs) of the
$\Delta m_{ni}$ matrix.  The five PCs contributing the most variance are
plotted in the top row of Figure~\ref{snc3}. The first three of these are seen to
vary smoothly across the focal plane and are clearly related to the patterns seen in
the drift solutions for the star flats (top row of
Figure~\ref{drifts}).  The lower row of Figure~\ref{snc3} plots the
temporal behavior of these PCs: their amplitudes are seen to change
slowly over the course of each observing season.  PC1 is a roughly
north-south gradient that changes by several mmag across the FOV over
the lifetime of the camera. 

PCs beyond the third exhibit little spatial coherence and lack overall
temporal trends.  We find that PC4 correlates with the seeing, and
suspect that it maps small errors in our nonlinearity corrections for
the amplifiers.  This effect is small, producing, $\ll1$~mmag RMS
photometric variation for bright stars.  Further PCs are even less
significant.

We conclude that the camera's photometric response changes by up to
$\pm7$~mmag on scales of months.  Any changes on scales of days are
well below 1~mmag.  The response variations are low-order functions of
focal-plane position and hence likely to be caused by slow
accumulations of dust or contaminants on the optical surfaces.  

The few-mmag shifts in CCD response seen in $Y$ band are thought to be
an exception, since the changes in CCD operating temperatures occur
when the focal plane is cycled to room temperature and back, a few
times per year.  The SN fields are not observed in $Y$ band so we
cannot verify this hypothesis directly.


\section{Implications}
\label{summary}

The photometric characterization of $\approx 1800$ specialized DECam
images of rich stellar fields yields a response model that is accurate
at mmag level across the focal plane and across 4 years of DECam
operations.  The ability to homogenize response of the instrument and
(on cloudless nights) the atmosphere to this level encourages future
attempts to calibrate the entire DES survey at accuracy significantly
better than the $\approx 7$~mmag RMS repeatability achieved for DES by
\citet{fgcm} and for PS1/SDSS by \citet{hypercal}.  A successful
few-mmag calibration model for DECam requires these key elements:
\begin{itemize}
\item Stable camera and electronics, such that the response of the
  system on a clear night stays constant over a period longer than
  the time required to characterize it.
\item Combination of internal-consistency constraints (``ubercal'')
  with prior assumptions on atmospheric and instrumental stability to
  remove large-scale degeneracies.
\item An instrumental reference response map $\vrr(t)$ that is free of
  the spurious signals present in dome flats due to stray light,
  pixel-scale variations, and spectral mismatches between the
  flat-field illumination and stellar spectra.
\item Accommodation of spatial and temporal variations of the spectral
  response $\vrr(\lambda,t)$ from the reference ``natural'' passband
  of the instrument, which can be empirically determined in the form
  of a color correction $\vc(t)$.
\item Use of a measurable aperture correction proxy, such as the fraction of
  the PSF found in a large but finite annulus, to compensate for
  temporal variation in atmospheric scattering of light out of the
  photometric aperture.
\item Allowance for slow (weeks to months), low-order drifts in the
  spatial structure of $\vrr(t)$.  These drifts are close to, but not
  quite, wavelength independent.  The camera response is much more
  stable than the dome screen illumination.
\item Recognition of the contribution of sky-estimation errors to the
  uncertainty budget for photometry.
\item Achieving mmag repeatability in $Y$ band would additionally
  require a spatial correction for flat-field errors related to the
  mounting structure of the CCDs, and a temporal correction for shifts
  in CCD operating temperatures.
\end{itemize}

After implementing these techniques, we find that the $griz$
stellar photometry of the star-flat exposures is highly repeatable.
In most exposures on most nights, the key quantifications of this are:
\begin{itemize}
\item The level of correlated photometric errors, as measured by the
  correlation function $\xi_m(\theta)$ of stellar magnitude residuals
  on scales $20\arcsec<\theta<80\arcsec,$ is 1~mmag$^2$ or lower.
\item The RMS magnitude errors, in excess of those expected from shot
  noise and read noise, are 2--3~mmag.  Effects that can be making
  this (zero-lag) RMS larger than the
  $\approx 1$~mmag of correlated error include: sky estimation errors;
  small-scale variations in pixel size due to lithography ``noise'';
  limitations of the approximation of bandpass variation by a linear
  color term; or, on longer time scales, stellar variability.
\item After application of the aperture correction proxy terms, the
  RMS deviation of exposure zeropoints from a simple atmospheric secant
  law is $\approx 1$~mmag.
\item A wavelength-independent 4th-order polynomial function of focal
  plane coordinates captures temporal changes in array response to
  $<1$~mmag accuracy in $i$ and $z$ bands, with up to 2~mmag RMS
  variation in $g$ and $r$ bands.  The blue component of the response
  drift could also be modeled to improve $g$ and $r$ accuracy.
\end{itemize}
DECam is remarkably stable and well-behaved.  For example, no changes
in amplifier gain have been detected (with the exception of a 3\% shift
in the gain of one amplifier of CCD S30 when it came back to life
after being non-functional for 3 years).  The spectral response shows
no evidence of change over time aside from that expected from variability
of atmospheric constituents.  The spatial response does change over
time by up to $\pm7$~mmag (Figure~\ref{drifts}, top row).  Although
we do not know the physical origin of this drift, it is gradual and
hence calibratable.

The transmission of the atmosphere on cloudless nights at airmass
$X<2$ is found to be
described by a secant law to 1 part per thousand on most nights.
Instances where correlated photometric errors rise perceptibly above 1~mmag
are all found to be correlated with
the aperture-correction proxy $A_t$, so we can conclude that the
primary photometric impact of atmospheric perturbations is via
scattering of light outside our 6\arcsec\ aperture.  The aperture
corrections can change rapidly and unpredictably by up to $\approx
10$~mmag, and some nights show elevated variability.  Fortunately
the 
6\arcsec--8\arcsec\ annular flux fraction ($A_t$) is directly measurable from
the images and can be used to correct these temporal, and perhaps
spatial, variations in aperture correction.

Our star-flat observations span only 25~minutes of clock time in each
filter.  Further investigation will be necessary to characterize the
amount of variation in atmospheric transmission that occurs over the
course of an entire evening.  This will be an important aspect of
extending our achieved level of accuracy to a global photometric
solution for DES.   The DES observing strategy is well-suited to the
task of global homogenization of photometry: each spot on the sky will
be observed in a given filter 8--10 times, on different parts of the
array, on different nights in different years, averaging down any remnant
spatial or temporal errors in photometric calibration.  Exposures are heavily
interlaced for strong internal constraints, and widely separated areas
of sky are observed on each cloudless night to generate constraints on
large-scale modes of the calibration solution.

These DECam results should be indicative of the techniques and
potential for other ground-based surveys to move from $\approx10$~mmag
photometric calibration toward the $\approx1$~mmag regime.  In
particular, the Large Synoptic Survey Telescope (LSST) will be able to
employ all of the methods that succeed with DECam, with vastly more
exposures per sky location and stellar detections per exposure, albeit
with the substantial additional complication of calibrating dependence
upon the varying camera/instrument rotation angle required for LSST's
alt-az mount.  Our
study emphasizes the utility of periodic ``star flat'' observing
sequences, with a range of dither steps on a single field, and also
the calibration utility of more frequent observations of fixed pointings
such as the supernova search areas.  Attempts to push photometric
calibration of ground-based visible imaging to $<1$~mmag accuracy will
need to address several new problems, including: characterization of
pixel-to-pixel lithographic variation; the inadequacy of simple linear
color corrections in describing the departures of system throughput
from the natural bandpass; inaccuracies in our aperture-correction
proxy; and, ultimately, scintillation.  Specialized observing
procedures, such as defocussing, have been used to overcome these
barriers, but may not be practical for general-purpose survey
observations.  But averaging of many repeat observations, with
dithered exposures, will serve to beat down most of these potential
sub-mmag systematic errors.

\acknowledgements
GMB gratefully acknowledges support from grants AST-1311924 and AST-1615555
from the National Science Foundation, and DE-SC0007901 from the Department
of Energy. 
Funding for the DES Projects has been
provided by the U.S. Department of Energy, the U.S. National Science
Foundation, the Ministry of Science and Education of Spain, the
Science and Technology Facilities Council of the United Kingdom, the
Higher Education Funding Council for England, the National Center for
Supercomputing Applications at the University of Illinois at
Urbana-Champaign, the Kavli Institute of Cosmological Physics at the
University of Chicago, the Center for Cosmology and Astro-Particle
Physics at the Ohio State University, the Mitchell Institute for
Fundamental Physics and Astronomy at Texas A\&M University,
Financiadora de Estudos e Projetos, Funda{\c c}{\~a}o Carlos Chagas
Filho de Amparo {\`a} Pesquisa do Estado do Rio de Janeiro, Conselho
Nacional de Desenvolvimento Cient{\'i}fico e Tecnol{\'o}gico and the
Minist{\'e}rio da Ci{\^e}ncia, Tecnologia e Inova{\c c}{\~a}o, the
Deutsche Forschungsgemeinschaft and the Collaborating Institutions in
the Dark Energy Survey.

The Collaborating Institutions are Argonne National Laboratory, the
University of California at Santa Cruz, the University of Cambridge,
Centro de Investigaciones Energ{\'e}ticas, Medioambientales y
Tecnol{\'o}gicas-Madrid, the University of Chicago, University College
London, the DES-Brazil Consortium, the University of Edinburgh, the
Eidgen{\"o}ssische Technische Hochschule (ETH) Z{\"u}rich, Fermi
National Accelerator Laboratory, the University of Illinois at
Urbana-Champaign, the Institut de Ci{\`e}ncies de l'Espai (IEEC/CSIC),
the Institut de F{\'i}sica d'Altes Energies, Lawrence Berkeley
National Laboratory, the Ludwig-Maximilians Universit{\"a}t
M{\"u}nchen and the associated Excellence Cluster Universe, the
University of Michigan, the National Optical Astronomy Observatory,
the University of Nottingham, The Ohio State University, the
University of Pennsylvania, the University of Portsmouth, SLAC
National Accelerator Laboratory, Stanford University, the University
of Sussex, Texas A\&M University, and the OzDES Membership Consortium.

Based in part on observations at Cerro Tololo Inter-American
Observatory, National Optical Astronomy Observatory, which is operated
by the Association of Universities for Research in Astronomy (AURA)
under a cooperative agreement with the National Science Foundation.

The DES data management system is supported by the National Science
Foundation under Grant Numbers AST-1138766 and AST-1536171.  The DES
participants from Spanish institutions are partially supported by
MINECO under grants AYA2015-71825, ESP2015-66861, FPA2015-68048,
SEV-2016-0588, SEV-2016-0597, and MDM-2015-0509, some of which include
ERDF funds from the European Union. IFAE is partially funded by the
CERCA program of the Generalitat de Catalunya.  Research leading to
these results has received funding from the European Research Council
under the European Union's Seventh Framework Program (FP7/2007-2013)
including ERC grant agreements 240672, 291329, and 306478.  We
acknowledge support from the Australian Research Council Centre of
Excellence for All-sky Astrophysics (CAASTRO), through project number
CE110001020.

This manuscript has been authored by Fermi Research Alliance, LLC
under Contract No. DE-AC02-07CH11359 with the U.S. Department of
Energy, Office of Science, Office of High Energy Physics. The United
States Government retains and the publisher, by accepting the article
for publication, acknowledges that the United States Government
retains a non-exclusive, paid-up, irrevocable, world-wide license to
publish or reproduce the published form of this manuscript, or allow
others to do so, for United States Government purposes.


\newpage
\begin{thebibliography}{}

\bibitem[Antilogus \etal(2014)]{antilogus}
Antilogus, P., Astier, P., Doherty, P., Guyonnet, A., \& Regnault, N.\ 2014, Journal of Instrumentation, 9, C03048

\bibitem[Bernstein \etal(2017a)]{decamast}
Bernstein,  G.~M., Armstrong, R., Plazas, A.~A., \etal\ 2017, \pasp, 129,  074503
 
\bibitem[Bernstein \etal(2017b)]{detrend}
Bernstein, G. \etal, 2017b, arxiv:1706.09928

\bibitem[Bertin(2006)]{scamp} 
Bertin, E.\ 2006, Astronomical Data Analysis Software and Systems XV, 351, 112 

\bibitem[Bertin \& Arnouts(1996)]{sextractor} 
Bertin, E., \& Arnouts, S.\ 1996, \aaps, 117, 393 

\bibitem[Betoule \etal(2014)]{betoule} 
Betoule, M., Kessler, R., Guy, J., \etal\ 2014, \aap, 568, A22 

\bibitem[Burke \etal(2014)]{Burke}
Burke, D. L. \etal\ (2014), \aj, 147, 19

\bibitem[Burke \etal(2017)]{fgcm}
Burke, E. \etal\ (2017), arxiv:1706.01542

\bibitem[Drlica-Wagner \etal(2017)]{y1gold} 
Drlica-Wagner, A., Sevilla-Noarbe, I., Rykoff, E.~S., \etal\ 2017, arXiv:1708.01531 

\bibitem[Estrada et al.(2010)]{estrada} 
Estrada, J., Alvarez, R., Abbott, T., \etal\ 2010, \procspie, 7735, 77351R 

\bibitem[Finkbeiner \etal(2016)]{hypercal} 
Finkbeiner, D.~P., Schlafly, E.~F., Schlegel, D.~J., \etal\ 2016,
\apj, 822, 66 

\bibitem[Flaugher \etal(2015)]{decam} 
Flaugher, B., Diehl, H.~T., Honscheid, K., \etal\ 2015, \aj, 150, 150 

\bibitem[Henden \& Munari(2014)]{apass} 
Henden, A., \& Munari, U.\ 2014, Contributions of the Astronomical Observatory Skalnate Pleso, 43, 518 

\bibitem[Li \etal(2016)]{ting}
Li, T.~S., DePoy, D.~L., Marshall, J.~L., \etal\ 2016, \aj, 151, 157 

\bibitem[Plazas \etal(2014)]{andres}
Plazas, A.~A., Bernstein, G.~M., \& Sheldon, E.~S.\ 2014, \pasp, 126, 750 

\bibitem[Gruen \etal(2015)]{gruen} 
Gruen, D., Bernstein, G.~M., Jarvis, M., \etal\ 2015, Journal of Instrumentation, 10, C05032 

\bibitem[Gunn \& Westphal(1981)]{GunnWestphal} 
Gunn, J.~E., \& Westphal, J.~A.\ 1981, \procspie, 290, 16 

\bibitem[Jacoby \etal(1998)]{mosaic} 
Jacoby, G.~H., Liang, M., Vaughnn, D., Reed, R., \& Armandroff, T.\ 1998, \procspie, 3355, 721 

\bibitem[Manfroid(1995)]{manfroid} 
Manfroid, J.\ 1995, \aaps, 113, 587 

\bibitem[McLeod \etal(1995)]{brian} 
McLeod, B.~A., Bernstein, G.~M., Rieke, M.~J., Tollestrup, E.~V., 
\& Fazio, G.~G.\ 1995, \apjs, 96, 117 

\bibitem[Morganson \etal(2017)]{desdm}
Morganson, E. \etal\ (2017), in preparation

\bibitem[Padmanabhan \etal(2008)]{nikhil}
Padmanabhan, N., Schlegel, D.~J., Finkbeiner, D.~P., \etal\ 2008, \apj, 674, 1217-1233

\bibitem[Press \etal(2003)]{recipes}
Press, W. H., Teukolsky, S. A., Vetterling, W. T., \& Flannery,
B. P. 2003, Numerical Recipes in C++, Cambridge University Press

\bibitem[Reil \etal(2014)]{rasicam} 
Reil, K., Lewis, P., Schindler, R., \& Zhang, Z.\ 2014, \procspie, 9149, 91490U 

\bibitem[Skrutskie \etal(2006)]{twomass} 
Skrutskie, M.~F., Cutri, R.~M., Stiening, R., \etal\ 2006, \aj, 131, 1163 

\bibitem[Tyson(1986)]{Tyson86} Tyson, J.~A.\ 1986, Journal of 
the Optical Society of America A, 3, 2131 


\end{thebibliography}
\end{document}